\documentstyle[12pt]{article}

\newcommand{\matel}[3]{\langle #1|#2|#3\rangle}
\newcommand{\ra}{\rightarrow}

\newcommand{\sG}{\sigma \cdot G}

\newcommand{\aver}[1]{\langle #1\rangle}

\newcommand{\as}{\alpha_s}

\newcommand{\MeV}{\,\mbox{MeV}}

\newlength{\dinwidth}
\newlength{\dinmargin}
\begin{document}
{}~~\\
\begin{flushright}
UND-HEP-95-BIG02\\
June 1995\\
Version 3.0\\
\end{flushright}
\vspace{1cm}
\begin{center}
\begin{Large}
\begin{bf}
%
%
THE QCD PERSPECTIVE ON LIFETIMES OF HEAVY-FLAVOUR HADRONS
\footnote{Preliminary version; final version will form part of
a Physics Report.}\\
\end{bf}
\end{Large}
\vspace{5mm}
\begin{large}
%
%
I.I. Bigi\\
\end{large}
%
%
Physics Dept., University of Notre Dame du Lac\\
Notre Dame, IN 46556, U.S.A.\\
e-mail address: BIGI@UNDHEP.HEP.ND.EDU
\vspace{5mm}
\end{center}
\noindent
\begin{abstract}
Over the last few years a theoretical treatment for the
weak decays of
heavy-flavour hadrons has been developed that is genuinely
based on QCD. Its methodology is described as it applies to
total lifetimes, and the underlying theoretical issues are
discussed. Theoretical expectations are compared with present
data. One discrepancy emerges: the beauty baryon lifetime
appears to be significantly shorter than predicted. The
ramifications of those findings are analyzed in detail, and
future refinements are described.
\end{abstract}
%
%
\section{Introduction}
\subsection{Goals and Obstacles}
As explained in detail elsewhere in this
review, a precise measurement of the lifetimes of the
various weakly decaying charm
and beauty hadrons possesses great experimental
value per se as well as in searches for
$B^0 - \bar B^0$ and $D^0 - \bar D^0$ oscillations. Yet there exists
also a strong theoretical interest in determining and
interpreting those lifetimes;
I want to sketch that
first in rather qualitative terms and
give specifics later.

Weak decays of hadrons depend on fundamental parameters
of the Standard Model, in particular on the KM parameters and
quark masses. It is eminently important to reliably determine
their values from data.
Alas -- this is easier said than done theoretically (and
experimentally)! For in such an endeavour we have to face the
"Dichotomy of the Two Worlds". On the one hand there is the
"Theorists' World" where quarks and gluons are the relevant
strongly interacting entities;  it is in this short-distance or
Femto World where theorists like to formulate their
fundamental
theories. On the other hand there is the "Real World" where
hadrons constitute the relevant degrees of freedom; it is in
that world where everyone (including theorists) lives and
measurements are performed. To formulate predictions from
the Theorists' World in the language of the Real World and to
translate findings from the Real World back into the idioms of
the Theorists' World represents the theoretical challenge one
faces.

One quantitative measure for the difference between the two
worlds is provided by the lifetimes of the weakly decaying
hadrons carrying the same flavour. On the quark level
there is obviously only a single lifetime for a given flavour.
Yet in the `Real
World' hadrons carrying the same flavour
quantum number possess different
(and even vastly different) lifetimes; e.g.,
$\tau (K^+)/\tau (K_S)\simeq 140$,
$\tau (D^+)/\tau (D^0)\simeq 2.5$ and
$0.9 \leq \tau (B^-)/\tau (B_d)\leq 1.2$. On the other hand
deviations of the
lifetime ratios from unity evidently decrease for
an
increasing heavy-flavour mass. This is as expected
in a simple `two-component' picture:
in the limit of
$m_Q$ -- the mass of the heavy-flavour quark $Q$ --
going to infinity
the `Spectator Ansatz' should hold
where the lifetimes of
all hadrons $H_Q$ containing $Q$ coincide; for finite
values of $m_Q$ there are {\em pre-asymptotic} corrections.
Two lines of reasoning support this qualitative picture:

\noindent ($\alpha$) The decay width of a quark $Q$ increases very
quickly
with
its mass $m_Q$:
$$\Gamma _Q \propto G_F^2 m_Q^5 \eqno(1)$$
for $m_Q \ll M_W$ changing into
$\Gamma _Q \propto \alpha m_Q^3/M_W^2$ for
$m_Q \gg M_W$ \footnote{The asymptotic scaling behaviour
$\Gamma _Q \propto m_Q^3$ (rather than
$\Gamma _Q\propto m_Q$) reflects
the coupling of the {\em longitudinal} $W$ boson -- the original
Higgs field -- to the top quark.}. Therefore for $m_Q$ sufficiently
large, its decay width is bound to exceed $\Lambda _{QCD}$,
i.e. the quark $Q$ decays before it can hadronize
\footnote{For top decays this happens for
$m_t\geq 130$ GeV \cite{RAPALLO}.}. Then there is
obviously a universal lifetime for such a heavy flavour and
the spectator ansatz applies trivially.

\noindent ($\beta$) Analysing all
{\em non}-spectator reactions explicitely, one finds their widths to
increase
with a smaller power of $m_Q$ than the spectator process;
thus one arrives
at the spectator ansatz as the asymptotic case.

The most relevant phenomenological question then is how quickly
the limit of universal lifetimes is approached.

\subsection{Phenomenology: Legends with Truths}
Some mechanisms had
been identified very early on that generate differences
in the lifetimes of hadrons $H_Q$ with the same heavy
flavour $Q$ : "Weak Annihilation" (=WA) of $Q$ with the light
valence antiquark for mesons or
"W Scattering" (=WS) with the valence diquark system for
baryons \footnote{A distinction is often made between W exchange
in the
s and in the t channel with the former case referred to
as `weak annihilation' and the latter as `W exchange'. This
classification
is however artificial since the two operators mix already under
one-loop renormalization in QCD, as discussed later on.
Both cases will summarily be
referred to as WA.}. Such an analysis had first been undertaken
for charm decays.

Since WA contributes to Cabibbo allowed decays of $D^0$, but not of
$D^+$
mesons (in the valence quark description), it creates a difference in
$\tau (D^0)$ vs. $\tau (D^+)$. However the WA rate is doubly
suppressed relative to the spectator rate, namely by the helicity
factor $(m_q/m_c)^2$ with
$m_q$ denoting the largest mass in the final state and by the
`wavefunction overlap' factor $(f_D/m_c)^2$
reflecting the practically zero
range of the low-energy weak interactions:
$$\Gamma ^{(0)}_{W-X}(D^0)
\propto G_F^2f^2_Dm_q^2m_c \, .\eqno(2)$$
Therefore it had
originally been suggested that already charm hadrons should
possess approximately equal lifetimes. It then came
as quite a
surprise when observations showed it to be otherwise -- in
particular since the first data suggested a considerably larger
value for $\tau (D^+)/\tau (D^0)$  than measured today. This
near-shock caused a re-appraisal of the theoretical situation; its
results at that time can be summarized in
four main points:

\noindent (i)  One source for a lifetime difference had
too quickly been discarded as insignificant. Cabibbo-allowed
nonleptonic decays of $D^+$
-- but not of $D^0$ -- mesons
produce two antiquarks in the final state that
carry the same flavour:
$$D^+ = [c\bar d] \ra (s\bar d u) \bar d $$
Thus one has to allow for the {\em interference}
between different
quark diagrams in $D^+$ , yet not in $D^0$ decays; the
$\bar d$ valence antiquark in $D^+$ mesons thus ceases to
play the role of an uninvolved bystander and a difference
in the $D^+$ vs. $D^0$ lifetimes will arise.
While this interference had been included in descriptions of
the $D\ra K \pi$ two-body modes it was
ignored for total widths. For it was
thought (often without stating it
explicitely) that the required coherence
would not be maintained between the two amplitudes
when applied to inclusive transitions.
This assumption was challenged
in ref.\cite{PI} where it was argued that even the two
inclusive amplitudes remain
sufficiently coherent. The interference
turns out to be destructive, i.e. it prolongs $\tau (D^+)$
over $\tau (D^0)$, but only once the QCD radiative corrections
have been included. This effect is usually referred to as
`Pauli Interference' (=PI) although such a name would be
misleading if it is interpreted as suggesting that the interference
is automatically destructive.

\noindent (ii) It was argued \cite{SONI} that the helicity suppression
of the WA contribution to $D$ decays can be vitiated. The diagram in
Fig.1 contains gluon bremsstrahlung off the
initial antiquark line in the $W$-exchange reaction;
evaluating this particular diagram
explicitely one finds:
$$\Gamma ^{(1)}_{W-X}(D^0)\propto
(\as /\pi ) G_F^2(f_D/\aver{E_{\bar q}})^2m_c^5
\eqno(3)$$
with
$\aver {E_{\bar q}}$ denoting the average energy of the
initial antiquark $\bar q$
\footnote{The $1/\aver {E_{\bar q}}$ term in the amplitude
derives from the
propagator in the diagram.}. Using a non-relativistic
wavefunction for the decaying meson one has
$\aver{E_{\bar q}}\simeq m_q$. This
contribution, although of higher order in $\as$, would
dominate over the lowest order term
$\Gamma ^{(0)}_{W-X}$ since helicity suppression has
apparently been vitiated and the decay constant $f_D$
is now
calibrated by $\aver{E_{\bar q}}$ with
$f_D/\aver{E_{\bar q}} \sim {\cal O}(1)$ rather than
$f_D/m_c \ll 1$.
The spectator picture would still apply at asymptotic quark
masses, since $\Gamma ^{(1)}_{W-X}/
\Gamma _c\propto (f_D/\aver{E_{\bar q}})^2 \ra 0$
as $m_c\ra \infty$ due to $f_D \propto 1/\sqrt{m_c}$.
Yet
if eq.(3) were indeed to hold, it would have a
dramatic impact on the theoretical description of weak
heavy-flavour decays: the impact of this
particular pre-asymptotic
correction, namely WA, would be enhanced considerably and
actually be quite significant even in beauty decays.
Alternatively
it had been suggested \cite{MINKOWSKI}
that the wavefunction of the $D$ meson
contains a $c \bar q g$ component where the $c \bar q$ pair forms
a spin-one configuration with the gluon $g$ balancing the spin
of the $c\bar q$ pair.

\noindent Both effects, namely PI and WA, work in the same
direction, i.e. both enhance $\tau (D^+)$ over $\tau (D^0)$.

\noindent (iii) A rich structure emerges in the decays of
charm baryons \cite{BARYONS1,BARYONS2,BARYONS3}:

$\bullet$  On the one hand WS contributes to the Cabibbo
allowed $\Lambda _c$ and $\Xi _c^0$ decays; one should
also keep in
mind that WS is {\em not} helicity suppressed in baryon
decays already to lowest order in the strong coupling. It is still
reduced in size by the corresponding wavefunction overlap;
yet that is at least partially off-set by WS being described by
two-body phasespace versus the three-body phase space of the
spectator transition; this relative enhancement in
phase space can be
estimated to be roughly of order $16\pi ^2$.

$\bullet$  PI affects the
$\Lambda _c$, $\Xi _c^{0,+}$ and $\Omega _c$
widths in various
ways, generating destructive as well as constructive contributions!
This also strongly suggests that it is very hard to make
reliable numerical predictions for these baryonic lifetimes; yet
the overall qualitative pattern has been predicted:
$$\tau (\Xi _c ^0) < \tau (\Lambda _c) < \tau (\Xi _c ^+) \eqno(4a)$$
together with
$$\tau (\Lambda _c)< \tau (D^0) < \tau (D^+)
\; . \eqno(4b)$$

\noindent (iv) Pre-asymptotic corrections might be sizeable in the
lifetime ratios of beauty hadrons.

\noindent Reviews of these phenomenological descriptions
can be found in \cite{RUCKL,BRADLEE}.

It turned out, as discussed in more detail later on, that some
of the phenomenological descriptions anticipated the
correct results: it is PI that provides the main
engine behind
the $D^+$-$D^0$ lifetime ratio; $\Lambda _c$ is considerably
shorter-lived than $D^0$; the observed charm baryon
lifetimes do obey the hierarchy stated in eq.(4).

Nevertheless the phenomenological treatments had significant
shortcomings, both of a theoretical and of a phenomenological
nature: (i) No agreement had emerged in the literature about how
corrections in particular due to WA and WS scale with the heavy
quark mass $m_Q$. (ii) Accordingly no clear predictions could be
made on the lifetime ratios among beauty hadrons, namely whether
$\tau (B^+)$ and $\tau (B_d)$ differ by a few to several percent
only, or by 20 - 30 \%, or by even more! (iii) No unequivocal
prediction on
$\tau (D_s)$ or $\tau (B_s)$ had appeared. (iv) In the absence of a
systematic
treatment it is easy to overlook relevant contributions, and that is
actually what happened; or the absence of certain corrections had to
be postulated in an ad-hoc fashion. Thus there existed an
intellectual as well as practical need for a description based
on a systematic theoretical framework rather than a set of
phenomenological prescriptions.

\subsection{From Phenomenology to Theory}
In the last few years we have succeeded
in showing that the non-perturbative corrections to heavy-flavour
decays can be expressed
through a {\em systematic} expansion in {\em inverse} powers of
$m_Q$. A simple analogy with nuclear $\beta$ decay can
illustrate this point. There are two effects distinguishing the decays
of neutrons bound in a nucleus from the decay of free neutrons:

\noindent (a) nuclear binding effects;

\noindent (b) the impact of Pauli statistics correlating the electrons
surrounding
the nucleus with those emerging from $\beta$ decay.

\noindent The typical energies of the bound electrons
-- $\epsilon _{el}$ -- are
certainly small compared to $E_{release}$, the energy released in the
decay; let us assume -- although this is not true in reality -- that also
the nuclear binding energies $\epsilon _{nucl}$ were small compared
to $E_{release}$. In that case nuclear $\beta$ decays would proceed,
to a good approximation, like the decays of {\em free} neutrons;
corrections to this simple `spectator' picture could be computed via
an expansion in powers of $\epsilon _{nucl}/E_{release}$ and
$\epsilon _{el}/E_{release}$. In practice, however, the corrections for
nuclear $\beta$ decay are incorporated by explicitely using the wave
functions of the bound nucleons and electrons (obtained with the
help of some fairly massive computer codes).

There arise analogous corrections to the decay rate for a quark $Q$
inside a hadron $H_Q$:

\noindent (a) interactions of the decaying quark with other partons
in the
hadron; this includes WA of
$Q$ with the light
valence antiquark for mesons or WS
with the valence
diquark system for baryons; they correspond to K
capture of bound electrons by a heavy nucleus in the preceding
example.

\noindent (b)  PI effects of the decay products with other
partons in the hadron; e.g.: $b\bar u \ra c\bar u d \bar u$
or $c\bar d \ra u \bar d s \bar d$.
They prolong the lifetimes of $D^+$ and $B^-$ mesons.

\noindent The difference to the example of nuclear $\beta$ decay
is quite
obvious: even in the limit $m_Q\ra \infty$ a non-relativistic
bound-state treatment is inapplicable since the dynamical
degrees of
freedom of the heavy-flavour hadron $H_Q$ cannot fully be
described by a hadronic wavefunction.  The most reliable approach is
then to evaluate weak decay rates of heavy-flavour hadrons through
an expansion in powers of $\mu _{had}/m_Q$ where
$\mu _{had}$ represents a hadronic scale $\leq$ 1 GeV.
The first few terms in this series should yield a good approximation
for beauty decays;
the situation for charm decays is a priori unclear
(and at present remains so a
posteriori as well); this will be discussed in detail later on.
\footnote{As already mentioned, top quarks decay weakly
before they can hadronize \cite{RAPALLO}.}
The vice of hadronization is then transformed into (almost) a virtue:
the weak decays of heavy-flavour hadrons constitute an intriguing
and novel laboratory for
studying strong dynamics through their
interplay with the weak forces -- and this is the secondary
motivation for studying them.
To be more specific: the heavy-flavour mass $m_Q$ provides an
expansion parameter that allows to deal with the
non-perturbative dynamics of QCD in a novel way.
The formalism to be employed
combines the $1/m_Q$ expansion with other elements derived from
QCD proper without having to invoke a `deus ex machina' --
in contrast to phenomenological descriptions.

There is one concept underlying, in one form or another, all efforts to
deal with hadronization, namely the notion of quark-hadron duality
(henceforth referred to as {\em duality} for short). In its broadest
formulation it can be stated as follows: sufficiently inclusive
transition rates  between hadronic systems can be calculated in
terms of quarks and gluons. While the general validity of
this concept has
not been established in a rigorous fashion, new light has been shed
on its nature, validity and applicability by $1/m_Q$ expansions.
Lifetimes obviously represent the most inclusive quantity where
duality should be applicable.
Nonleptonic transitions are certainly more complex than semileptonic
ones; yet I will argue later that while there probably exists a
quantitative difference in the degree to which duality holds in
semileptonic and in nonleptonic decays, there is {\em no qualitative}
one. Deatiled measurements of
lifetimes are then theoretically important not only to translate data
on the semileptonic branching ratio into a determination of the
semileptonic
width, but also in their own right.

Dedicated and comprehensive studies of both charm and beauty
decays are called for. The KM parameters relevant for charm decays
-- $V(cs)$
and $V(cd)$ -- are well known through unitarity constraints of the
3x3 KM
matrix, in contrast to the situation in beauty decays, and
I consider it unlikely that charm decay studies can improve on that.
On the other hand those can be used to calibrate our theoretical
tools before applying them to beauty decays.

Before concluding this general introduction I want to point out a less
straightforward aspect of accurate lifetime measurements: decay rate
evolutions in proper time for neutral mesons will not follow a single
exponential function when particle-antiparticle oscillations occur.
For there exist two distinct mass eigenstates with
$\Delta m \equiv m_1-m_2\neq 0 \neq
\Delta \Gamma \equiv \Gamma _1-\Gamma _2 $. The quantity
$\Delta m$ generates a deviation of the form
$e^{-\Gamma t}\cos \Delta mt$
or $e^{-\Gamma t}\sin \Delta mt$;
$\Delta \Gamma \neq 0$
leads to the emergence of a second exponential.
The general expression reads as follows:
$$d\Gamma (B[D]\ra f)/dt \; \propto
e^{-\Gamma t}\cdot G(t)$$
$$G(t) = a+be^{-\Delta \Gamma t}+
ce^{-1/2\Delta \Gamma t}\cos \Delta mt +
de^{-1/2\Delta \Gamma t}\sin \Delta mt\eqno(5a)$$
where
$$a=|A(f)|^2\left[ \frac{1}{2}
(1+|\frac{q}{p}\bar \rho (f)|^2)
+ Re[ \frac{q}{p}\bar \rho (f)]
\right] $$
$$b=|A(f)|^2\left[ \frac{1}{2}
[ 1+|\frac{q}{p}\bar \rho (f)|^2]
- Re[ \frac{q}{p}\bar \rho (f)]
\right] $$
$$c=|A(f)|^2\{
1-|\frac{q}{p}\bar \rho (f)|^2\}\; , \;
d= 2|A(f)|^2 Im[ \frac{q}{p}\bar \rho (f)] \; , \;
\bar \rho (f)= \frac{\bar A(f)}{A(f)}
\eqno(5b) $$
with $\bar A(f)$ and $A(f)$ denoting the amplitude  for
$\bar B [\bar D] \ra f$ and $B [D] \ra f$, respectively.

\section{Preview of the Predictions on the
Lifetime Ratios for Beauty and Charm Hadrons}
In this section I summarize the numerical results and
sketch the main elements of the underlying theoretical
treatment in a way that can satisfy the casual reader. A more
in-depth discussion of the theoretical concepts and tools
will be presented in subsequent sections.

Expanding the width for the decay of a heavy-flavour hadron
$H_Q$ containing $Q$
into an inclusive final state $f$ through order $1/m_Q^3$
one obtains
\cite{BUV,BS,SV}
$$\Gamma (H_Q\ra f)=\frac{G_F^2m_Q^5}{192\pi ^3}|KM|^2
\left[ c_3^f\matel{H_Q}{\bar QQ}{H_Q}_{norm}+
c_5^f\frac{
\matel{H_Q}{\bar Qi \sG Q}{H_Q}_{norm}}{m_Q^2}+ \right.
$$
$$\left. +\sum _i c_{6,i}^f\frac{\matel{H_Q}
{(\bar Q\Gamma _iq)(\bar q\Gamma _iQ)}{H_Q}_{norm}}
{m_Q^3} + {\cal O}(1/m_Q^4)\right] ,  \eqno(6)$$
where the dimensionless coefficients $c_i^f$ depend on the
parton level
characteristics of $f$ (such as the ratios of the final-state quark
masses
to $m_Q$); $KM$ denotes the appropriate combination of KM
parameters,
and $\sG = \sigma _{\mu \nu}G_{\mu \nu}$
with $G_{\mu \nu}$ being the gluonic field strength tensor. The last
term
in eq.(6)
implies also the summation over the four-fermion operators with
different light flavours $q$. The expectation values of the local
operators appearing on the right-hand side of eq.(6)
contain the
relativistic normalization of the state $|H_Q\rangle$:
$$\matel{H_Q}{O_i}{H_Q}_{norm} \equiv
\matel{H_Q}{O_i}{H_Q}/2M_{H_Q} \, .\eqno(7) $$

It is through the quantities
$\matel{H_Q}{O_i}{H_Q}$ that the dependence on the {\em decaying
hadron} $H_Q$, and
on
non-perturbative forces in general, enters, instead of through
wavefunctions as in nuclear $\beta$ decay.
Since these are matrix
elements
for on-shell hadrons $H_Q$, one sees that $\Gamma (H_Q\ra f)$ is
indeed
expanded into a power series in $\mu _{had}/m_Q < 1$. For
$m_Q\ra \infty$ the contribution from the lowest dimensional
operator obviously dominates; here it is the dimension three
operator $\bar QQ$. A heavy quark expansion yields:
$$\matel{H_Q}{\bar QQ}{H_Q}_{norm}=1+
{\cal O}(1/m_Q^2)\, , \eqno(8)$$
i.e. $\matel{H_Q}{\bar QQ}{H_Q}_{norm}=1$ for $m_Q\ra \infty$,
reflecting the unit of heavy-flavour common to all
hadrons $H_Q$.

Eqs.(6,8) show that the leading nonperturbative corrections are
of order $1/m_Q^2$ rather than $1/m_Q$; therefore they can be
expected to be rather small in beauty decays. The expectation
values appearing in the $1/m_Q^2$ and
$1/m_Q^3$ contributions can be related to other observables
and their size thus be extracted in a reliable
manner. Applying the
master formula eq.(6) to lifetimes one arrives at the following
general results:

\noindent $\bullet$ The leading contribution to the total decay
width is given by the first term on the
right-hand-side of eq.(8) that is
{\em common} to all hadrons
of a given heavy-flavour quantum number.
For $m_Q\ra \infty$ one has thus derived -- from QCD proper -- the
spectator picture attributing equal lifetimes to all hadrons of a
given heavy-flavour! This is not a surprising result -- after all
without hadronization there is of course only a unique lifetime --,
but it is gratifying nevertheless.

\noindent $\bullet$ Lifetime differences first arise at order
$1/m_Q^2$ and are controlled by the expectation
values of two dimension five operators, to be discussed later. These
terms, which had
been overlooked in the original phenomenological analyses, generate
a lifetime difference between heavy-flavour {\em baryons} on one
side  and {\em mesons} on the other. Yet apart from
small isospin or $SU(3)_{fl}$ breaking they shift the
meson widths by the same amount and thus do {\em not}
lead to differences among the meson lifetimes.

\noindent $\bullet$ Those emerge at
order $1/m_Q^3$ and are expressed through the expectation values
of four-fermion operators which are proportional to
$f_M^2$ with $f_M$ denoting the decay constant for the meson
$M$.  Contributions from what is referred to as
WA and PI in the
original phenomenological descriptions
(see the discussion in Sect.1) are systematically and
consistently included. Further
contributions to the baryon-meson lifetime difference also
arise at this level due to WS.

\noindent $\bullet$ Since the transitions $b\ra c l \nu$ or
$c\ra s l \nu$ are described by an isosinglet operator one can
invoke the isospin invariance of the strong interactions to
deduce for the semileptonic widths
$$ \Gamma _{SL}(B^-)= \Gamma _{SL}(B_d) \eqno(9a)$$
$$ \Gamma _{SL}(D^+)= \Gamma _{SL}(D^0) \eqno(9b)$$
and therefore
$$\frac{\tau (B^-)}{\tau (B_d)}=
\frac{BR_{SL}(B^-)}{BR_{SL}(B_d)} \eqno(10a)$$
$$\frac{\tau (D^+)}{\tau (D^0)}=
\frac{BR_{SL}(D^+)}{BR_{SL}(D^0)} \eqno(10b)$$
up to small corrections due to the KM [Cabibbo]
suppressed transition $b\ra u l \nu$
[$c\ra d l \nu$] which changes isospin by half a unit.
The spectator picture goes well beyond eqs.(9): it
assigns the same semileptonic width to all hadrons of
a given heavy flavour. Yet such a property cannot be
deduced on {\em general} grounds: for one had to rely
on $SU(3)_{Fl}$ symmetry to relate
$\Gamma _{SL}(D_s)$ to $\Gamma _{SL}(D^0)$ or
$\Gamma _{SL}(B_s)$ to $\Gamma _{SL}(B_d)$ and
no symmetry can be invoked to relate the semileptonic
widths of mesons and baryons. There is actually a WA
process that generates semileptonic decays on the
Cabibbo-allowed level for $D_s$ [and also for
$B_c$], but not for $D^0$ and $D^+$ nor for
$B_d$, $B^-$ and $B_s$ mesons: the hadrons are produced
by gluon emission off the $\bar s$ [or the $\bar c$] line.
Yet since the relative weight of WA is significantly reduced
in meson decays, one does not expect this mechanism to
change $\Gamma _{SL}(D_s)$ significantly relative to
$\Gamma _{SL}(D^0)$. Actually there are corrections
to the semileptonic
widths arising already in order $1/m_Q^2$. On rather
general grounds one predicts the expectation values
$\matel{P_Q}{\bar QQ}{P_Q}$ and
$\matel{P_Q}{\bar Qi\sG Q}{P_Q}$ to be largely
independant of the flavour of the light antiquark in the
meson and therefore
$$\Gamma _{SL}(D_s)\simeq \Gamma _{SL}(D^0) \eqno(11a)$$
$$\Gamma _{SL}(B_s)\simeq \Gamma _{SL}(B_d) \eqno(11b)$$
like in the naive spectator picture, but for
non-trivial reasons. On the other hand,
as explained later, the values of the expectation values
for these operators are different when taken between
baryon states and one expects
$$\Gamma _{SL}(\Lambda _Q) > \Gamma _{SL}(P_Q)
\eqno(11c)$$

The remarks above can be summarized as follows:
$$\Gamma (\Lambda _Q) = \Gamma ([Q\bar q]^0) =
\Gamma ([Q\bar q]^{\pm}) + {\cal O}(1/m_Q^2) \eqno(12a)$$
$$\Gamma (\Lambda _Q) > \Gamma ([Q\bar q]^0) =
\Gamma ([Q\bar q]^{\pm}) + {\cal O}(1/m_Q^3) \eqno(12b)$$
$$\Gamma (\Lambda _Q) > \Gamma ([Q\bar q]^0) >
\Gamma ([Q\bar q]^{\pm}) + {\cal O}(1/m_Q^4) \eqno(12c)$$

Quantitative predictions for the lifetime ratios of beauty
hadrons through order $1/m_b^3$ are given in Table \ref{TABLE1}
together with present data.
\begin{table}
\begin{tabular} {|l|l|l|}
\hline
Observable &QCD ($1/m_b$ expansion) &Data \\
\hline
\hline
$\tau (B^-)/\tau (B_d)$ & $1+
0.05(f_B/200\, \MeV )^2
[1\pm {\cal O}(10\%)]>1$ &$1.04 \pm 0.05$ \\
&(mainly due to {\em destructive} interference) & \\
\hline
$\bar \tau (B_s)/\tau (B_d)$ &$1\pm {\cal O}(0.01)$
&  $ 0.98\pm 0.08$ \\
\hline
$\tau (\Lambda _b)/\tau (B_d)$&$\sim 0.9\; ^*$ & $0.76\pm 0.06$
\\
\hline
\end{tabular}
\centering
\caption{QCD Predictions for Beauty Lifetimes}
\label{TABLE1}
\end{table}
The expectations \cite{MIRAGE,MARBELLA,DS}
for the lifetimes of charm hadrons are juxtaposed to  the data in
Table \ref{TABLE2}.
The overall agreement is rather good given the theoretical and
experimental uncertainties. It has to be kept in mind that the
{\em numerical} predictions on {\em baryon} lifetimes involve
quark model results for the size of the relevant expectation
values. This is indicated in Tables 1 and 2 by the asterisk.
There is however one obvious discrepancy: the observed
$\Lambda _b$ lifetimes is significantly shorter than predicted.
While -- as indicated just above -- predictions on {\em baryon}
lifetimes have to be taken with a grain of salt, one cannot change
the prediction on
$\tau (\Lambda _b)/\tau (B_d)$ with
complete theoretical impunity; this will be explained later on.
\begin{table}
\begin{tabular} {|l|l|l|}
\hline
Observable &QCD ($1/m_c$ expansion) &Data \\
\hline
\hline
$\tau (D^+)/\tau (D^0)$ & $\sim 2 \; \; \; $
[for $f_D \simeq 200$ MeV] &$2.547 \pm 0.043$ \\
&(mainly due to {\em destructive} interference) & \\
\hline
$\tau (D_s)/\tau (D^0)$ &$1\pm$ few $\times 0.01$
&  $ 1.125\pm 0.042$ \\
\hline
$\tau (\Lambda _c)/\tau (D^0)$&$\sim 0.5\; ^*$ &
$0.51\pm 0.05$\\
\hline
$\tau (\Xi ^+ _c)/\tau (\Lambda _c)$&$\sim 1.3\; ^*$ &
$1.75\pm 0.36$\\
\hline
$\tau (\Xi ^+ _c)/\tau (\Xi ^0 _c)$&$\sim 2.8\; ^*$ &
$3.57\pm 0.91$\\
\hline
$\tau (\Xi ^+ _c)/\tau (\Omega _c)$&$\sim 4\; ^* $&
$3.9 \pm 1.7$\\
\hline
\end{tabular}
\centering
\caption{QCD Predictions for Charm Lifetimes}
\label{TABLE2}
\end{table}

These numerical predictions bear out the general expectations
expressed above. This
will be discussed below in more detail.

For proper perspective on the role played by the heavy quark
expansion some comments should be made at this point:

\noindent (i) As stated before, the early phenomenological
treatments had already identified the relevant mechanisms by
which lifetime differences arise, namely PI, WA and WS.
Furthermore the authors of ref.\cite{PI} had argued
-- correctly -- that PI constitutes the dominant effect. Yet it
is the heavy quark expansion that provides a firm
theoretical underpinning
to these expectations. In particular it clarifies -- both
conceptually and quantitatively -- the role of WA and
how its weight scales with $1/m_Q$.

 \noindent (ii) Contributions of order $1/m_Q$ would
dominate all other effects -- if they were present! The heavy
quark expansion shows unequivocally that they are absent in
total rates; this has important ramifications,  to be pointed out
later.

\noindent (iii) Corrections of order $1/m_Q^2$ differentiate
between the decays of mesons and of baryons. They had been
overlooked before. The heavy quark expansion has identified them
and basically determined their size.

\section{Methodology of the Heavy Quark Expansion for Fully
Integrated Rates}
The weak decay of the heavy quark $Q$ inside the heavy-flavour
hadron $H_Q$
proceeds within a cloud of light degrees of freedom (quarks,
antiquarks
and gluons) with which $Q$ and its decay products can interact
strongly. It
is the challenge for theorists to treat these initial and final state
hadronization effects. Among the existing four Post-Voodoo
theoretical technologies -- QCD Sum Rules,
Lattice QCD, Heavy Quark Effective Theory (HQET) and Heavy Quark
Expansions -- only the last one deals with inclusive decays. On
the other hand it benefits, as
we will see, from the results of the other three
technologies, since those can determine the size of some of
the relevant expectation values.\footnote{It should be noted that
contrary
to frequent claims in the literature HQET -- as it is usually
and properly defined -- per se does {\em not} allow to treat
lifetimes or even the total semileptonic width:
for those observables strongly depend on $m_Q$ whereas it
is the special feature of HQET that $m_Q$ is removed from
its Lagrangian.}

In analogy to the treatment of
$e^+e^-\rightarrow hadrons$ one describes the transition rate into an
inclusive final state $f$ through the imaginary part of a
forward scattering operator evaluated to second order in the weak
interactions \cite{BUV,BS,SV}:
$$\hat T(Q\rightarrow f\rightarrow Q)=
i \, Im\, \int d^4x\{ {\cal L}_W(x){\cal L}_W^{\dagger}(0)\}
_T\eqno(13)$$
where $\{ .\} _T$ denotes the time ordered product and
${\cal L}_W$ the relevant effective weak Lagrangian expressed on
the
parton level. If the energy released in the decay is sufficiently large
one can express the {\em non-local} operator product in eq.(13) as an
infinite sum of {\em local} operators $O_i$ of increasing dimension
with
coefficients $\tilde c_i$
containing higher and higher powers of
$1/m_Q$.\footnote{It should be kept in mind,
though, that
it is primarily the {\em energy release} rather than $m_Q$ that controls the
expansion.} The width for $H_Q\rightarrow f$ is then obtained by
taking the
expectation value of $\hat T$ between the state $H_Q$:
$$\matel{H_Q}{\hat T (Q\ra f\ra Q)}{H_Q} \propto
\Gamma (H_Q\ra f) = G_F^2 |KM|^2
\sum _i \tilde c_i^{(f)} \matel{H_Q}{O_i}{H_Q}
\eqno(14)$$
This master formula holds for a host of different inclusive
heavy-flavour decays: semileptonic, nonleptonic and radiative
transitions,
KM favoured or suppressed etc. For semileptonic and nonleptonic
decays treated through order $1/m_Q^3$ it takes the form already
given in eq. (6):
$$\Gamma (H_Q\ra f)=\frac{G_F^2m_Q^5}{192\pi ^3}|KM|^2
\left[ c_3^f\matel{H_Q}{\bar QQ}{H_Q}_{norm}+
c_5^f\frac{
\matel{H_Q}{\bar Qi\sG Q}{H_Q}_{norm}}{m_Q^2}+ \right.
$$
$$\left. +\sum _i c_{6,i}^f\frac{\matel{H_Q}
{(\bar Q\Gamma _iq)(\bar q\Gamma _iQ)}{H_Q}_{norm}}
{m_Q^3} + {\cal O}(1/m_Q^4)\right] $$
Integrating out the two internal loops in Figs.2 and 3
yields the operators $\bar QQ$ and $\bar Qi\sG Q$,
respectively; the
black boxes represent ${\cal L_W}$.
Likewise the diagrams in Figs.4 generate four-quark
operators; Fig.4a and 4b differ in how the light quark flavours
are connected inside the hadron $H_Q$:
cutting a quark line $q$ in Fig.2 and connecting it to the $\bar q$
constituent of the $H_Q$ meson, as shown in Fig.4a, one has a WA
transition operator; cutting instead the
$\bar q$ line and connecting it to the $H_Q$ constituents, see
Fig.4b, leads to the four-fermion operator describing PI.
Inspection of these diagrams suggests a rule-of-thumb
that is borne out by explicit computations: the dimensionless
coefficients $c^f_d$ are smaller for the two-loop diagrams of
Figs.2 and 3 than for the one-loop diagrams of Figs.4.

Integrating out the internal loops of Fig.3 actually generates
not only the dimension-five chromomagnetic operator, but also
dimension-six quark-gluon operators,
namely $\bar Q (iD_{\mu}G_{\mu \nu})\gamma _{\nu}Q$ and
$\bar Q\sigma _{\mu \nu}G_{\mu \rho}\gamma _{\rho}
iD_{\nu}Q$. Using the equations of motion those can be
reduced to four-quark operators. Yet their coefficients are
suppressed by $\alpha _S/\pi$ relative to the coefficient
coming from Figs.4; therefore one can discard their contributions
at the present level of accuracy.

One important general observation should be made at this
point. The two operators $\bar QQ$ and $\bar Qi\sG Q$ obviously
do not contain any light quark or antiquark fields; those enter
first on the dimension-six level. Their expectation values are,
as discussed below, practically insensitive to the light degrees
of freedom up to terms of order $1/m_Q^3$. The
inclusive semileptonic and
non-leptonic widths through order $1/m_Q^2$ are thus controled
by the {\em decay} of a single quark $Q$,
yet occurring in a non-trivial background field; the
{\em spectator} contribution universal to the widths for all
hadrons $H_Q$ of a given heavy flavour constitutes its
asymptotic part:
$$\Gamma (H_Q) \simeq \Gamma _{decay}(H_Q) +
{\cal O}(1/m_Q^3)\; , \eqno(15a)$$
$$\Gamma _{decay}(H_Q) \simeq \Gamma _{spect}(Q) +
{\cal O}(1/m_Q^2)\; . \eqno(15b)$$
However $\Gamma _{decay}$ possesses a different value
for mesons and baryons, as discussed below.

The expansion introduced above will be of practical use only
if its first few terms provide a good approximation and if
one can determine the corresponding expectation values
in a reliable manner. The latter can be achieved in two
complementary ways: One relates the matrix element in
question to other observables
through a heavy quark expansion; or one harnesses
other second-generation theoretical
technologies, namely QCD sum rules, QCD
simulations on the lattice and Heavy Quark Effective Theory,
to calculate these quantities. The results obtained so far
are listed below:

(a) Using the equations of motion one finds for the leading operator
$\bar QQ$:
$$\bar QQ= \bar Q\gamma _0Q -
\frac{\bar Q[(i\vec D)^2-(i/2)\sG]Q}{2m_Q^2}+
g_S^2\frac{\bar Q\gamma _0t^iQ
\sum _q \bar q\gamma _0 t^iq}{4m_Q^3} +
{\cal O}(1/m_Q^4)\eqno(16)$$
with the sum in the last term running over the light quarks $q$; the
$t^i$ denote the
colour $SU(3)$ generators. I have ignored total derivatives in this
expansion since they do not contribute to the $H_Q$ expectation
values.  There emerge two dimension-five operators, namely
$\bar Q (i\vec D)^2 Q$ and $\bar Q \sG Q$ with $\vec D$
denoting the covariant derivative. The first one represents the
square of the spatial momentum of the heavy quark $Q$ moving in
the soft gluon background and thus describes its kinetic
energy\footnote{Since it is not a Lorentz scalar, it cannot
appear in
eq.(6).}. The second one constitutes the chromomagnetic operator
that already appeared in eq.(6). The
relevant dimension-six operators can be expressed as four-fermion
operators. Since $\bar Q\gamma _0 Q$ constitutes the Noether
current for the heavy-flavour quantum number one has
$\matel{H_Q}{\bar Q\gamma _0Q}{H_Q}_{norm}=1$ leading to eq.(8).

(b)  The first non-perturbative corrections arise at
order $1/m_Q^2$, and they can enter in two ways, namely
through the chromomagnetic operator in the OPE of $\hat T$ or
through the $1/m_Q$ expansion of
the $H_Q$
expectation values of the local operator $\bar QQ$.

(c) For the chromomagnetic operator one finds
$$\matel{P_Q}{\bar Qi\sG Q}{P_Q}_{norm} \simeq
\frac{3}{2} (M_{V_Q}^2-M_{P_Q}^2) \eqno(17a)$$
where $P_Q$ and $V_Q$ denote the pseudoscalar and vector mesons,
respectively.  For $\Lambda _Q$ and $\Xi _Q$ baryons one has
instead
$$\matel{\Lambda _Q}{\bar Qi\sG Q}{\Lambda _Q}
\simeq 0 \simeq \matel{\Xi _Q}{\bar Qi\sG Q}{\Xi _Q}\eqno(17b)$$
since the light diquark system inside $\Lambda _Q$ and $\Xi _Q$
carries no spin. These baryons thus represent (though only through
order $1/m_Q^2$) a simpler system than the mesons where the light
antiquark of course carries spin. For $\Omega _Q$ baryons on the
other hand the situation is different: for the light di-quarks
$ss$ form a spin-one configuration. The $\Omega _Q$
expectation value is then given by the spin-splitting in the
baryon masses:
$$\matel{\Omega _Q}{\bar Qi\sG Q}{\Omega _Q}_{norm}
\simeq \frac {4}{3} [M^2(\Omega _Q^{(3/2)}) -
M^2(\Omega _Q)]\eqno(17c)$$

(d) We thus have
$$\matel{\Lambda _Q}{\bar QQ}{\Lambda _Q}_{norm}=1 -
\frac{\langle (\vec p_Q)^2\rangle _{\Lambda _Q}}{2m_Q^2}+
{\cal O}(1/m_Q^3)\eqno(18a)$$
$$\matel{P_Q}{\bar QQ}{P_Q}_{norm}=1 -
\frac{\langle (\vec p_Q)^2\rangle _{P_Q}}{2m_Q^2}+
\frac{3}{8} \frac{M_{V_Q}^2-M_{P_Q}^2}{m_Q^2}+
{\cal O}(1/m_Q^3)\eqno(18b)$$
with the notation $\langle (\vec p_Q)^2\rangle _{H_Q}
\equiv \matel {H_Q}{\bar Q (i\vec D)^2 Q}{H_Q}_{norm}$.
The reason for the appearance of the kinetic energy term
in eqs. (18) is quite transparent: The first two terms on the r.h.s. of
eqs.(18) represent the mean value of the factor
$\sqrt{1-\vec v^2}$ reflecting the time dilation slowing down the
decay of the quark $Q$ moving inside $H_Q$.

(e) The value of $\langle (\vec p_Q)^2\rangle _{H_Q}$
has not been determined yet in an accurate way
(details will be given later);  we know, though, it
has to obey the inequality \cite{VOLOSHIN,OPTICAL}
$$\langle (\vec p_Q)^2\rangle _{H_Q}\geq \frac{1}{2}
\matel{H_Q}{\bar Qi\sG Q}{H_Q}_{norm}
\equiv \aver{\mu _G^2}_{P_Q}\eqno(19)$$
which -- because of eq.(17) -- is useful only for $H_Q=P_Q$.
\footnote{One should note that this inequality has now been derived
in a fully field-theoretical treatment of QCD, rather than a merely
quantum-mechanical one. Its proper interpretation
will be discussed later.}
On the other hand, the difference in the kinetic energy of Q inside
baryons and mesons can be related to the masses of charm and
beauty hadrons \cite{BUVPREPRINT}:
$$ \langle (\vec p_Q)^2\rangle _{\Lambda _Q}-
\langle (\vec p_Q)^2\rangle _{P_Q} \simeq
\frac{2m_bm_c}{m_b-m_c}\cdot
\{ [\langle M_B\rangle -M_{\Lambda _b}]-
[\langle M_D\rangle -M_{\Lambda _c}] \} \eqno(20)$$
with $\langle M_{B,D}\rangle$ denoting the `spin averaged' meson
masses:
$$ \langle M_B\rangle \equiv \frac{1}{4}(M_B+3M_{B^*})
\eqno(21)$$
and likewise for $\langle M_D\rangle$. In deriving eq.(20) it was
implicitely assumed that
the $c$ quark can be treated also as heavy; in that case
$\langle (\vec p_c)^2\rangle _{H_c} \simeq
\langle (\vec p_b)^2\rangle _{H_b}$ holds.

(f) Eqs.(17-18,20) show that the two
{\em dimension-five} operators do
produce differences in $P_Q$ vs. $\Lambda _Q/\Xi _Q$ vs.
$\Omega _Q$ lifetimes
(and semileptonic widths) of order
$1/m_Q^2$; i.e., $\Gamma _{decay}(H_Q)$, as defined in
eqs.(15), depends on whether $H_Q$ represents a (pseudoscalar)
meson or a baryon. In the latter case it is
sensitive to whether the light diquark system carries spin
zero ($\Lambda _Q$, $\Xi _Q$) or
one ($\Omega _Q$). The origin of such a difference is quite transparent.
(i) The heavy quark $Q$ can be expected to
possess a somewhat different kinetic
energy inside a meson or a baryon; it can also be affected by spin-spin
interactions with the light di-quarks.
This difference in the heavy quark motion means that
Lorentz time dilation prolongs the lifetime of the quark $Q$
to different degrees in baryons than in mesons.
(ii) While there exists a spin interaction
between $Q$ and the antiquark $\bar q$ in the meson or the
$ss$ system in the $\Omega _Q$ baryon,
there is no such coupling inside $\Lambda _Q$
or $\Xi _Q$.

\noindent Yet these effects do not generate a significant difference
among
{\em meson} lifetimes since their expectation values satisfy isospin
and $SU(3)_{fl}$ symmetry to a good degree of accuracy.

(g) Differences in meson lifetimes are generated at order
$1/m_Q^3$ by {\em dimension-six} four-quark operators
describing PI as well as WA; the weight of the latter is, as
explained in more detail in the next section, greatly reduced.
The expectation values of these operators
look very similar to the one controlling $B^0-\bar B^0$ oscillations:
$$\matel{H_Q(p)}{(\bar Q_L\gamma _{\mu}q_L)
(\bar q_L\gamma _{\nu}Q_L}{H_Q(p)}_{norm}\simeq
\frac{1}{4m_{H_Q}}f^2_{H_Q}p_{\mu}p_{\nu}\eqno(22)$$
with $f_{H_Q}$ denoting the decay constant for the meson $H_Q$;
the so-called bag factor has been set to unity, i.e.
{\em factorization}
has been assumed.

(h) The situation becomes much more complex for $\Lambda _Q$ and
baryon decays in general. To order $1/m_Q^3$ baryons lose the
relative simplicity mentioned above: there are several different
ways in which the valence quarks of the baryon can be contracted
with the quark fields in the four-quark operators; furthermore
WS is {\em not} helicity suppressed and thus can make a sizeable
contribution to lifetime differences; also the PI effects can
now be constructive
as well as destructive. Finally one cannot take
recourse to factorisation as a limiting case.
Thus there emerge three types of
numerically significant mechanisms at this order in baryon
decays -- in contrast to meson decays where there is a
{\em single}  dominant
source for lifetime differences -- and their strength cannot be
expressed in terms of
a single observable like $f_{H_Q}$. At present we do not know
how to determine the relevant matrix elements in a
model-independant way. The best available guidance and
inspiration is to be gleaned from quark model calculations with
their inherent uncertainties. This analysis had already been
undertaken
in the framework of phenomenological models
\cite{BARYONS1,BARYONS2,BARYONS3} with the following
qualitative results:

\noindent -- WS contributes to $\Lambda _Q$ and $\Xi _Q^0$
decays;

\noindent -- a {\em destructive} interference affects
$\Lambda _Q$, $\Xi _Q$ and $\Omega _b$ transitions.

\noindent -- a {\em constructive} interference enhances
$\Xi _c$ and $\Omega _c$  decays.

\noindent One thing should be obvious already at this point:
with terms of different signs and somewhat uncertain size
contributing to differences among baryon lifetimes one has to
take even semi-quantitative predictions with
a grain of salt!

\noindent The probability amplitude for WS or interference to
occur is expressed in quark models by the wavefunction for
$Q$ and one of the light quarks at zero spatial separation. This
quantity is then related to the meson wavefunction through the
hyperfine splitting in the baryon and meson masses
\cite{BARYONS3,MARBELLA}:
$$\frac{|\psi _{\Lambda _Q}^{(Qq)}(0)|}{|\psi _{P_Q}(0)|}\simeq
\frac{2m_q^*(M_{\Sigma _Q}-M_{\Lambda _Q})}
{(M_{P_Q}-m_q^*)(M_{V_Q}-M_{P_Q})}\eqno(23)$$
with $m_q^*$ denoting the {\em phenomenological constituent}
mass
of the light quark $q$ to be employed in these models. For the
baryonic expectation values one then obtains
$$\matel{\Lambda _Q}{\bar Q\Gamma _{\mu}Q
\bar q \Gamma _{\mu} q}{\Lambda _Q}_{norm} \sim
\frac{1}{4\aver{\mu ^2_G}_{P_Q}}(M_{\Sigma _Q}-M_{\Lambda _Q})
m^*_q F^2_{P_Q}M_{P_Q}\eqno(24)$$
and likewise for $\Xi _Q$.
For the $\Omega _Q$ one has instead
$$\matel{\Omega _Q}{\bar Q\Gamma _{\mu}Q
\bar q \Gamma _{\mu} q}{\Omega _Q}_{norm} \sim
\frac{5}{6\aver{\mu ^2_G}_{P_Q}}(M_{\Sigma _Q}-M_{\Lambda _Q})
m^*_q F^2_{P_Q}M_{P_Q} \; . \eqno(25)$$
reflecting the different spin substructure which already
had caused the difference in eqs.(17a) vs. (17b).
For simplicity one has assumed in these expressions that
the hyperfine splitting in the baryon masses are universal.

To summarize this discussion:

\noindent $\bullet$ The non-perturbative corrections
to total widths through order $1/m_Q^3$ are expressed in
terms of three non-trivial (types of) matrix elements:
$\aver{\mu _G^2}_{H_Q}$, $\aver{(\vec p_Q)^2}_{H_Q}$ and
$\matel{H_Q}{(\bar Q\Gamma _i q)(\bar q\Gamma _i Q)}{H_Q}$.

\noindent $\bullet$ The size of $\aver{\mu _G^2}_{H_Q}$ is
well-known for the mesons and $\Lambda _Q$ and $\Xi _Q$
baryons; reasonable estimates can be obtained for
$\Omega _Q$ baryons.

\noindent $\bullet$ A lower bound exists for the mesonic
expectation value $\aver{(\vec p_Q)^2}_{P_Q}$, but not
for the baryonic one. Various arguments strongly suggest
that $\aver{(\vec p_Q)^2}_{H_Q}$ is close to its lower
bound. Furthermore the quantity
$\aver{(\vec p_Q)^2}_{\Lambda _Q}-\aver{(\vec p_Q)^2}_{P_Q}$,
which is highly relevant for the difference in baryon
vs. meson lifetimes (and semileptonic widths) can be extracted
from mass measurements.

\noindent $\bullet$ The expectation value of the
four-quark operators between {\em mesons} can be expressed
in terms of a single quantity, namely the decay
constant $f_{P_Q}$.

\noindent $\bullet$ However the {\em baryonic} expectation
values of the four-quark operators constitute a veritable
Pandora's box, which at present is beyond theoretical
control. On the other hand measurements of inclusive
weak decay rates and the resulting
understanding of the role played by WS and PI present us
with a novel probe of the internal structure of the
heavy-flavour baryons.

\section{Comments on the Underlying Concepts}
Before explaining these
predictions in more detail and commenting on the comparison with
the data in the next subsection, I want to add  four general remarks
and elaborate on them. While I hope they will elucidate the
underlying concepts, they are not truly essential for following the
subsequent discussion and the more casual reader can ignore them.

(A) {\em On the validity of the $1/m_Q$ expansion in the presence of
gluon emission:} The WA contribution to the decay width of
pseudoscalar mesons to lowest order in the strong coupling has
been sketched in eq.(2).
The diagram in Fig.1 contains gluon bremsstrahlung off the
initial antiquark line in the $W$-exchange reaction;
evaluating this diagram
explicitely one finds, as stated before:
$$\Gamma ^{(1)}_{W-X}\propto
(\as /\pi ) G_F^2(f_{H_Q}/\aver{E_{\bar q}})^2m_Q^5
\eqno(26)$$
with
$\aver {E_{\bar q}}$ denoting the average energy of the
initial antiquark $\bar q$.
Since $\Gamma ^{(1)}_{W-X} \propto 1/\aver{E_{\bar q}}^2$,
it  depends very sensitively on the
low-energy quantity $ \aver{E_{\bar q}}$, and
its magnitude is thus quite
uncertain. The presence of a $1/\aver{E_{\bar q}}^2$
(or $1/m_q^2$) term would also mean that the nonperturbative
corrections to even inclusive decay widths of heavy-flavour
hadrons are not `infrared safe' and cannot be treated consistently through an
expansion in inverse powers of the heavy quark mass
$m_Q$ only.

\noindent Fortunately the contribution stated in eq.(26)
turns out to be
spurious for {\em inclusive} transitions.
This can best be seen by
studying the imaginary part of the forward scattering
amplitude $Q\bar q \ra f \ra Q\bar q$ as shown in Fig.5.
There are actually three poles in this amplitude indicated
by the broken lines
which
represent different final states: one with an on-shell
gluon and the other two with an off-shell gluon
materializing as a $q\bar q$ pair and involving interference
with the spectator decay amplitude. These processes are
all of the same order in the strong coupling and therefore
have to be summed over for an {\em inclusive} decay.
Analysing carefully the analyticity properties of the sum of these
forward
scattering amplitudes one finds \cite{MIRAGE,WA}
that it remains finite in the limit
of $\aver{E_{\bar q}} \ra 0$, i.e. the amplitude for the
{\em inclusive} width does {\em not} contain terms of order
$1/\aver{E_{\bar q}}^2$ or even $1/\aver{E_{\bar q}}$! The
contribution of WA can then be described in terms of
the expectation value of a {\em local} four-fermion
operator although the final state is dominated by low-mass
hadronic systems.

\noindent To summarize: the quantity $f_{H_Q}$ is calibrated by the
large mass $m_Q$ rather
than by $\aver{E_{\bar q}}$;
WA is then only
moderately significant even in charm decays.
The $1/m_Q$ scaling thus persists to hold even in the presence of
radiative corrections -- but only for {\em inclusive} transitions!
This caveat is not of purely academic interest. For
it provides a nice toy model illustrating the {\em qualitative}
difference between inclusive and exclusive transitions. The latter are
represented here by the three separate cuts in Fig.5.
For charm decays they correspond to the reactions
$[c\bar u] \ra s \bar d g$ and $[c\bar u] \ra s \bar d u\bar u$.
The two individual transition rates quite sensitively depend on
a low energy scale - $\aver{E_{\bar q}}$ -
which considerably enhances the rate for the first
transition while reducing it for the second one.
Yet the dependance on $1/\aver{E_{\bar q}}$ disappears
from their sum!

(B) {\em On the fate of the corrections of order
$1/m_Q$:}
The most important element of eq.(2) is -- the one that
is
{\em missing}! Namely there is no term of order $1/m_Q$ in the
total decay width whereas such a correction definitely exists for the
mass formula: $M_{H_Q}=m_Q(1+\bar \Lambda /m_Q+{\cal
O}(1/m_Q^2))$ (and likewise for differential decay distributions).
Hadronization in the initial state does generate corrections
of order $1/m_Q$ to the total width, as does hadronization in the
final state. Yet
local colour symmetry enforces that they
cancel against each other!
\footnote{This can be nicely illustrated in quantum mechanical toy
models: The total rate for the transition $Q\ra q$ depends on the
{\em local} properties of the potential, i.e. on the potential around
$Q$ in a neighourhood of size
$1/m_Q$ only; the nature of the resulting spectrum for $q$
-- whether it is discrete or continuous, etc. -- depends
of course on the long range properties of the potential,
namely whether it is confining or not.}
This can be understood in
another more compact (though less intuitive) way as well: with the
leading operator $\bar QQ$ carrying dimension three only
dimension-four operators can generate
$1/m_Q$ corrections; yet there is no independent dimension-four
operator \cite{CHAY,BUV}
{\em once the equation of motion is imposed} -- unless one abandons
local
colour symmetry thus making the operators $\bar Qi\gamma \cdot
\partial Q$
and $\bar Qi\gamma \cdot B Q$ independent of each other
($B_{\mu}$ denotes the gluon field)! The leading
non-perturbative corrections to fully integrated decay widths are
then
of order $1/m_Q^2$ and their size is controlled by two
dimension-five operators, namely the chromomagnetic and the
kinetic
energy operators, as discussed above. Their contributions
amount to no more than 10 percent
for $B$ mesons -- $(\mu _{had}/m_b)^2$ $\simeq
{\cal O}\left( (1\, GeV/m_b)^2\right)
\sim {\cal O}(\% )$ -- as borne out by the detailed analysis
previewed in Table 1.

(C) {\em Which mass is it?}
For a $1/m_Q$ expansion it is of course important
to understand which kind of quark mass is to be employed there,
in particular since
for confined quarks there exists no a priori natural choice.
It had been claimed that the pole mass can and therefore
should conveniently be used. Yet such claims
turn out to be fallacious \cite{POLEMASS,BRAUN}: QCD, like QED, is
not
Borel summable; in the high order terms
of the perturbative series there arise instabilities
which are customarily referred to as
(infrared) renormalons representing poles in the Borel plane; they
lead to an {\em additive} mass renormalization generating an
{\em irreducible uncertainty}
of order $\bar \Lambda$ in the size of the pole mass:
$m_Q^{pole}=m_Q^{(0)}(1+c_1\as + c_2\as ^2 +...+ c_N\as ^N) +
{\cal O}(\bar \Lambda )=m_Q^{(N)}(1+{\cal O}(\bar \Lambda/
m_Q^{(N)}))$.   While this effect
can safely be ignored in a purely perturbative treatment, it negates
the
inclusion of non-perturbative corrections $\sim {\cal O}(1/m_Q^2)$,
since those are
then parametrically smaller than the uncertainty
$\sim {\cal O}(1/m_Q)$ in the definition of
the
pole mass. This problem can be taken care of through
Wilson's prescriptions for the operator product expansion:
$$\Gamma (H_Q\ra f)=\sum _i c_i^{(f)}(\mu )
\matel{H_Q}{O_i}{H_Q}_{(\mu )}\eqno(27)$$
where a momentum scale $\mu$ has been introduced to allow a
consistent
separation of contributions from Long Distance and Short Distance
dynamics -- $LD > \mu ^{-1} > SD$ -- with the latter contained in
the coefficients $c_i^{(f)}$ and the former lumped into the matrix
elements. The quantity $\mu$ obviously represents an auxiliary
variable which drops out from the observable, in this case the decay
width. In the limit
$\mu \ra 0$ infrared renormalons emerge in the coefficients; they
cancel
against ultraviolet renormalons in the matrix elements.
Yet that does
{\em not} mean that these infrared renormalons are irrelevant and
that
one can conveniently set $\mu =0$! For to incorporate both
perturbative
as well as non-perturbative corrections one has to steer a careful
course
between `Scylla' and `Charybdis': while one wants to pick
$\mu \ll m_Q$ so as to make a heavy quark expansion applicable,
one
also has to choose $\mu _{had}\ll \mu$ s.t. $\alpha _S(\mu ) \ll 1$;
for otherwise the {\em perturbative} corrections become
uncontrollable. Wilson's
OPE allows to incorporate both perturbative and non-perturbative
corrections, and {\em this underlies also a consistent formulation of
HQET};
the scale $\mu$ provides an infrared cut-off that automatically
freezes out infrared renormalons. For the asymptotic difference
between the hadron and the quark mass one then has to write
$\bar \Lambda (\mu ) \equiv
(M_{H_Q}-m_Q(\mu ))_{m_Q\ra \infty}$.
This nice feature does not come for free,
of course: for one has to use a `running' mass $m_Q(\mu )$
evaluated at an
intermediate scale $\mu$ which presents a technical
complication. On the other hand this quantity can reliably
be extracted from data \cite{OPTICAL}; furthermore it drops out
from
lifetime {\em ratios}, the main subject of this discussion.

(D) {\em On Duality:}
Quark-hadron duality equates transition rates calculated on
the quark-gluon level with the observable ones involving
the corresponding hadrons -- provided one sums over a
sufficient number of final states.
Since the early days of the quark model this concept has
been invoked in many different formulations; among other
things it has never been clearly defined what
constitutes a sufficient number of final states. This lack of
a precise formulation emerged since duality had never been
derived from QCD in a rigorous fashion; furthermore a certain
flexibility in an unproven, yet appealing intuitive concept can
be of considerable heuristic benefit.

\noindent Heavy quark expansions assume the validity of duality.
Nevertheless they provide new and fruitful insights into its
workings. As already stated, heavy quark expansions are based
on an OPE, see eq.(26), and as such are properly defined in Euclidean
space. There are two possible limitations to the procedure by which
duality is implemented in heavy quark expansions. (i) The
size of the coefficients $c_i^{(f)}$ is controlled by short-distance
dynamics. In concrete applications they are actually obtained
within perturbation theory. Those computations involve
integrals over all momenta. However for momenta below
the scale $\mu$ a perturbative treatment is unreliable.
The contributions from this regime might turn out to be
numerically insignificant for the problem at hand; yet in
any case one can undertake to incorporate them through the
expectation values of higher-dimensional operators -- the
so-called condensates. This procedure is the basis of what is
somestimes referred to as the `practical' version of the OPE.
Yet it is conceivable that there are significant short-distance
contributions from non-perturbative dynamics as well;
instantons provide an illustration for such a complication,
although their relevance is quite unclear at present \cite{MISHA}.
(ii) Once the OPE has been constructed and its coefficients
$c_i^f$ determined there, one employs a dispersion relation to
analytically continue them into Minkowski space. This means,
strictly speaking, that in Minskowski space only `smeared'
transition rates can be predicted, i.e. transition rates
averaged over some finite energy range. This situation was
first analyzed in evaluating the cross section for
$e^+e^- \ra had $ \cite{POGGIO}.
Through an OPE QCD allows to compute
$$\aver{\sigma (e^+e^- \ra had; E_{c.m.})}
\equiv \frac{1}{\Delta E_{sm}}
\int ^{E_{c.m.}+\Delta E_{sm}}_{E_{c.m.}-\Delta E_{sm}}
dE'_{c.m.} \sigma (e^+e^- \ra had; E'_{c.m.})$$
with
$0< \Delta E_{sm} \ll E_{c.m.}$. If the cross section happens to be
a smooth function of $E'_{c.m.}$ -- as it is the case well above
production thresholds --, then one can effectively take the limit
$\Delta E_{c.m.} \ra 0$ to predict
$\sigma (e^+e^- \ra had; E_{c.m.})$ for a
{\em fixed} c.m. energy $E_{c.m.}$. This scenario can be referred to
as {\em local} duality. Yet close to a threshold, like for charm or
beauty production, with its resonance structure one has to retain
$\Delta E_{sm} \sim \mu _{had}\sim 0.5 - 1$ GeV. The same
considerations are applied to heavy-flavour decays. Based on
{\em global} duality one can predict `smeared' decay rates, i.e.
decay rates averaged over a finite energy interval. If the energy
release is large enough the decay rate will be a smooth function of it,
smearing will no longer be required and {\em local} duality
emerges. How large the energy release has to be for this
simplification to occur cannot be predicted (yet). For the onset of
local duality is determined by terms that cannot be seen in any
finite order of the $1/m_Q$ expansion. However rather general
considerations lead to the following expectations: (i) This onset
should be quite abrupt around some energy scale exceeding usual
hadronic scales. (ii) Local duality should hold to a good degree of
accuracy for beauty decays, even for nonleptonic ones. (iii) On the
other hand there is reason for concern that this onset might not have
occurred for charm decays. This is sometimes referred to by saying
that inclusive charm decays might receive significant contributions
from (not-so-)`distant cuts'.

\section{Size of the Matrix Elements}

As discussed before, see eq.(27), the size of matrix elements depends
in general on the scale $\mu$ at which they are evaluated. This will
be addressed separately for the different cases.

The scalar dimension-three operator $\bar QQ$ can be expanded in
terms of $\bar Q\gamma _0Q$ -- with
$\matel{H_Q}{\bar Q\gamma _0Q}{H_Q}_{norm} = 1$ --
and operators of dimension five and higher, see eq.(15).

\noindent (i) Dimension-five operators

\noindent Employing eq.(16) for $b$ and $c$ quarks one finds:
$$\aver{\mu _G^2}_{B}\equiv
\matel{B}{\bar b\frac{i}{2}\sigma \cdot Gb}{B}_{norm} \simeq
\frac{3}{4} (M^2_{B^*}-M^2_B)\simeq 0.37\, (GeV)^2, \eqno(28a)$$
$$\aver{\mu _G^2}_D\equiv
\matel{D}{\bar c\frac{i}{2}\sigma \cdot Gc}{D}_{norm} \simeq
\frac{3}{4} (M^2_{D^*}-M^2_D)\simeq 0.41\, (GeV)^2,\eqno(28b)$$
i.e., these two matrix elements that have to coincide in the
infinite mass limit are already very close to each other.
They are already evaluated at the scale that is
appropriate for beauty and
charm decays.

For the description of $B_c$ decays one needs the expectation
value of the beauty as well as charm chromomagnetic operator
taken between the $B_c$ state. While the $B_c^*$ and $B_c$ masses
have not been measured yet, one expects the theoretical
predictions \cite{QUIGG} on them to be fairly reliable. With
$M(B^*_c)\simeq 6.33$ GeV and $M(B_c)\simeq 6.25$ GeV one obtains
$$\matel{B_c}{\bar b\frac{i}{2}\sigma \cdot Gb}{B_c}_{norm} \simeq
\matel{\bar B_c}{\bar c\frac{i}{2}\sigma \cdot Gc}{\bar B_c}_{norm}
\simeq 0.75\; (GeV)^2 \; ; \eqno(29)$$
i.e., double the value as for the mesons with light antiquarks.

{}From eq.(20) one obtains for the {\em difference}
in the kinetic energy
of the heavy quark inside baryons and mesons
$$\langle (\vec p_b)^2\rangle _{\Lambda _b}-
\langle (\vec p_b)^2\rangle _B
\simeq -(0.07 \pm 0.20)\, (GeV)^2 \eqno(30)$$
using available data; for charm the same value holds
to this order. We see that at present one cannot tell
whether the kinetic energy of the heavy quark
inside baryons exceeds that inside mesons or not. The
uncertainty in the mass of $\Lambda _b$ constitutes the bottleneck;
it would be quite desirable to decrease
its uncertainty down to below 10 MeV or even less. For
the overall size of $\langle (\vec p_Q)^2\rangle _{H_Q}$ we have
the following lower bounds from eq.(19):
$$\aver{(\vec p_b)^2}_B\geq 0.37 (GeV)^2\eqno(31)$$
$$\aver{(\vec p_c)^2}_D\geq 0.40 (GeV)^2\eqno(32)$$
To be conservative one can add a term of at most
$\pm 0.15\; (GeV)^2$ to the right-hand side of eqs.(31,32)
reflecting the uncertainty in the scale at which the
expectation values are to be evaluated
\cite{OPTICAL}. There are various
arguments as to why the size of $\aver{(\vec p_b)^2}_B$ will
not exceed this lower bound significantly. One should
note that such values for $\aver{(\vec p_b)^2}_B$ are
surprisingly large: for they state that the momenta of the
heavy quark inside the hadron is typically around 600 MeV.
An analysis based on QCD sum rules yields a value that is
consistent with the preceding discussion
\cite{QCDSR}:
$$\aver{(\vec p_b)^2}_B = (0.5\pm 0.1)\; (GeV)^2\eqno(33)$$
It can be expected that practically useful
values for $\aver{(\vec p_{c,b})^2}_{D,B}$ will be derived from
lattice QCD in the forseeable future
\cite{GUIDO}.

The leading non-perturbative corrections are thus controlled
by the following parameters:
$$\frac{\langle \mu ^2_G\rangle _B}{m_b^2}\simeq 0.016 \; \; ,
\frac {\langle (\vec p_b)^2\rangle _B}{m_b^2} \sim 0.016
\eqno(34a)$$
$$\frac{\langle \mu ^2_G\rangle _D}{m_c^2}\simeq 0.21 \; \; ,
\frac {\langle (\vec p_c)^2\rangle _D}{m_c^2} \sim 0.21
\eqno(34b)$$
$$\frac{\aver{\mu ^2_G}_{\Omega _c}}{m_c^2}
\simeq 0.12 \eqno(35)$$
The $\Omega _c$ mass hyperfine splitting has not been
measured yet. A quark model estimate of 70 MeV has
been used in evaluating eq.(17b).

\noindent (ii) Dimension-six Operators

\noindent The size of the decay
constants $f_B$, $f_{B_s}$, $f_D$ and
$f_{D_s}$ has not been determined yet with good accuracy.
Various theoretical technologies yield
$$f_D \sim 200\, \pm \, 30 \, MeV \eqno(36)$$
$$f_B \sim 180\, \pm \, 40 \, MeV \eqno(37)$$
$$\frac{f_{B_s}}{f_B} \simeq 1.1 - 1.2 \eqno(38)$$
$$\frac{f_{D_s}}{f_D} \simeq 1.1 - 1.2 \eqno(39)$$
where a higher reliability is attached to the predictions for the
ratios of decay constants, i.e. eqs.(38,39), than for their
absolute size, eqs.(36,37). Recent studies by WA75 and
by CLEO on $D_s^+\ra \mu ^+ \nu $ yielded values for $f_{D_s}$
that are somewhat larger, but still consistent with these predictions.
It can be expected that lattice QCD will produce
more precise results on these decay constants in the foreseeable
future and that those will be checked by future measurements
in a significant way. Yet there are two subtleties to be kept in mind
here:

\noindent ($\alpha$) The matrix element of the four-quark operator
is related to the decay constant through the assumption of
`vacuum saturation'. Such an ansatz cannot hold as an identity; it
represents an approximation the validity of which has to depend on
the scale at which the matrix element is evaluated. The quantities
$f_B$ and $f_D$ are measured in $B\ra \tau \nu ,\,  \mu \nu$ and
$D\ra \tau \nu ,\,  \mu \nu$, respectively and thus probed at the
heavy-flavour mass, $m_B$ and $m_D$. Yet for the strong
interactions controling the size of the expectation value of
the four-quark operator the heavy-flavour mass is a completely
foreign parameter. If vacuum saturation makes (approximate)
sense anywhere, it has to be at ordinary hadronic scales
$\mu \simeq 0.5 - 1$ GeV. That means the decay constant that
is observed at the heavy-flavour  scale has to be evaluated at the
hadronic scale $\mu$; this is achieved by the so-called hybrid
renormalization to be described later.

\noindent ($\beta$) The difference in decay widths generated by
the four-quark operators is stated as being proportional to
$f^2_{H_Q}/m_Q^2$ with
$f_{H_Q}= F_{H_Q}[1 - |\bar \mu |/m_Q +{\cal O}(1/m_Q^2)]$;
$F_{H_Q}$ represents the leading or `static' term which
behaves like $1/\sqrt{m_Q}$ for $m_Q\ra \infty$
and therefore $F_D > F_B$. Thus
asymptotically one has $F^2_{H_Q}/m_Q^2$ which indeed vanishes
like
$1/m_Q^3$. Yet the decay constant $f_{H_Q}$ -- partly as a
consequence of its
usual definition via an axialvector rather than a pseudoscalar
current -- contains large pre-asymptotic corrections which lead to
$f_D \sim f_B$. Then it is not clear which value to use
for $f_{H_Q}$ when calculating width differences, the asymptotic
one -- $F_{H_Q}$ -- or the one including the pre-asymptotic
corrections -- $f_{H_Q}$ where numerically
$F_{H_Q} > f_{H_Q}$ holds. .
It can be argued that the main impact of some of the
dimension-seven operators in the OPE for the meson decay width is
to renormalise the $F^2_{H_Q}/m_Q^2$ term
to $f^2_{H_Q}/m_Q^2$; yet this
issue can be decided only through computing the contributions from
dimension-seven operators. I will return to this issue when
discussing beauty and charm decays specifically.

Both these considerations also apply when determining the {\em
baryonic} expectation values of four-quark
operators \cite{MARBELLA}. Yet -- as already pointed out -- that is a
considerably more murky affair, as will become quite apparent
in the subsequent discussions of the beauty and charm lifetimes.

While there are significant uncertainties and ambiguities in the
values of the masses of beauty and charm quarks their difference
which is free of renormalon contributions is
tightly constrained:
$$m_b-m_c\simeq \langle M\rangle _B - \langle M\rangle _D +
\langle (\vec p)^2\rangle \cdot
\left( \frac{1}{2m_c}- \frac{1}{2m_b}\right)
\simeq 3.46 \pm 0.04\, GeV\, .\eqno(40)$$
This value agrees very well with the one extracted from
an analysis of energy spectra in semileptonic $B$ decays
\cite{VOLOSHIN2}.
Lifetime ratios are hardly sensitive to this difference; we list it here
mainly for completeness.

\section{Predictions on Beauty Lifetimes}

Within the Standard Model there is no real uncertainty
about the weak forces driving beauty
decays; at the scale $M_W$ they are given by the Lagrangian
$${\cal L}_W^{\Delta B=1}(\mu = M_W)=
\frac{4G_F}{\sqrt{2}}
[V_{cb}\bar c_L\gamma _{\mu}b_L+
V_{ub}\bar u_L\gamma _{\mu}b_L]\cdot
[V_{ud}^*\bar d_L\gamma _{\mu}u_L +
V_{cs}^*\bar s_L\gamma _{\mu}c_L]  \eqno(41)$$
for non-leptonic transitions with an analogous expression for
semileptonic ones. In eq.(41) I have ignored Cabibbo-suppressed
transitions and also the $b\ra t$ coupling since here we are not
interested in Penguin contributions. Radiative QCD corrections lead
to a well-known renormalization at scale $m_b$, which is often
referred to as {\em ultra-violet} renormalization:
$${\cal L}_W^{\Delta B=1}(\mu = m_b)=
\frac{4G_F}{\sqrt{2}} V_{cb}V_{ud}^*
\{ c_1(\bar c_L\gamma _{\mu}b_L)(\bar d_L\gamma _{\mu}u_L)
+c_2(\bar d_L\gamma _{\mu}b_L)(\bar c_L\gamma _{\mu}u_L)
\} \eqno(42)$$
for $b\ra c\bar ud$ transitions and an analogous expression
for $b\ra c \bar cs$;
the QCD corrections are lumped together into the coefficients $c_1$
and $c_2$ with
$$c_1=\frac{1}{2}(c_+ + c_-), \; \; c_2=\frac{1}{2}(c_+ - c_-)
\eqno(43a)$$
$$c_{\pm}=\left[ \frac{\alpha _S(M_W^2)}{\alpha _S(m_b^2)}
\right] ^{\gamma _{\pm}}, \; \;
\gamma _+=\frac{6}{33-2N_f}=-\frac{1}{2}\gamma _-
\eqno(43b)$$
in the leading-log approximation; $N_f$ denotes the number
of active flavours. Numerically this amounts to:
$$c_1(LL) \simeq 1.1, \; \; c_2(LL) \simeq -0.23 \, .
\eqno(44a)$$
Next-to-leading-log corrections modify mainly $c_2$
\cite{CHERNYAK}:
$$c_1(LL+NLL) \simeq 1.13, \; \; c_2(LL+NLL) \simeq -0.29 \, .
\eqno(44b)$$
Radiative QCD corrections thus lead to a mild enhancement of the
original coupling -- $c_1>1$ -- together with the appearance of an
induced operator with a different colour flow: $c_2\neq 0$. Later on
we will also include the so-called `hybrid' renormalization reflecting
radiative corrections in the domain from $m_b$ down to
$\mu _{had}$, the scale at which the hadronic expectation values
are to be evaluated.

Such perturbative corrections affect the overall scale of
the semileptonic and non-leptonic width (and
thus of the semileptonic branching ratio); but
by themselves they cannot generate {\em differences} among
the lifetimes of beauty hadrons. Those have to be
initiated by non-perturbative contributions although their
numerical size depends also on perturbative corrections.

The semileptonic and non-leptonic widths of
beauty {\em hadrons} $H_b$ through
order $1/m_b^2$ are given by
$$\frac{\Gamma _{SL,decay}(H_b)}{\Gamma _0} =
\matel{H_b}{\bar bb}{H_b}_{norm}\cdot
\left[ \eta _{SL}I_0(x,0,0) +
\frac{\aver{\mu _G^2}_{H_b}}{m_b^2}
(x\frac{d}{dx}-2)I_0(x,0,0)\right] \; , \eqno(45a)$$
$$\frac{\Gamma _{NL,decay}(H_b)}{\Gamma _0} =  N_C\cdot
\matel{H_b}{\bar bb}{H_b}_{norm}\cdot
\left[ \frac{A_0}{3}
[I_0(x,0,0)+ \rho _{c\bar c} I_0(x,x,0) +\right.
$$
$$\left. \frac{\aver{\mu _G^2}_{H_b}}{m_b^2}
(x\frac{d}{dx}-2)(I_0(x,0,0)+I_0(x,x,0))]  -
\frac{4}{3} A_2 \frac{\aver{\mu _G^2}_{H_b}}{m_b^2} \cdot
[I_2(x,0,0,) + I_2(x,x,0)] \right] \; , \eqno(45b)$$
$$\Gamma _0 \equiv
\frac{G_F^2m_b^5}{192 \pi ^3}|V(cb)|^2 \; . \eqno(45c)$$
The following notation has been used here:
$I_0$ and $I_2$ are phase-space factors, namely
$$I_0(x,0,0)= (1-x^2)(1-8x+x^2)-12x^2\log  x \eqno(46a)$$
$$I_2(x,0,0)=(1-x)^3\; , \; \; x=(m_c/m_b)^2 \eqno(46b)$$
for $b \ra c \bar ud/c l\bar \nu$ and
$$I_0(x,x,0)=v(1-14x-2x^2-12x^3) + 24x^2(1-x^2)
\log \frac{1+v}{1-v}\, , \, v=\sqrt{1-4x}\eqno(46c)$$
$$I_2(x,x,0)=v(1+\frac{x}{2}+3x^2)-3x(1-2x^2)
\log \frac{1+v}{1-v} \, , \eqno(46d)$$
for $b\ra c \bar cs$
transitions. The quantities $\eta _{SL}$, $\rho _{c\bar c}$,
$A_0$ and $A_2$ represent the
QCD radiative corrections. More specifically one has
$$A_2=(c_+^2-c_-^2)\; \; \; , \; \; \;
A_0= (c_-^2+2c_+^2)\cdot J \eqno(46e)$$
with $J$ reflecting the effect
of the subleading logarithms \cite{PETRARCA} and
$$\eta _{SL}\simeq 1+
\frac{2}{3}\frac{\as}{\pi}g(m_c/m_b,m_l/m_b,0)
\; . \eqno(46f)$$
The function $g$ can be computed numerically for arbitrary
arguments \cite{PHAM} and analytically for the most
interesting case $m_l=0$ \cite{NIR}.
The allowed values
for $\rho _{c\bar c}$,which reflects the fact that QCD
radiative corrections are quite sensitive to the final state quark
masses, can be found in ref.\cite{BAGAN}.

With $x\simeq 0.08$ one obtains
$$I_0(x,0,0)|_{x=0.08}\simeq 0.56 \; \; , \; \;
I_2(x,0,0)|_{x=0.08}\simeq 0.78   \; \; for \; \;
b\ra c \bar ud$$
$$I_0(x,x,0)|_{x=0.08}\simeq 0.24 \; \; , \; \;
I_2(x,x,0)|_{x=0.08}\simeq 0.32 \; \; for \; \;
b\ra c \bar cs$$
Since these functions are normalized to unity for $x=0$, one
notes that the final-state quark masses reduce the
available phase space quite considerably.

Some qualitative statements can illuminate the dynamical situation:

\noindent (i) As indicated in eqs.(15),
$\Gamma _{SL/NL,decay}$ differ from the naive
spectator result in order $1/m_b^2$.

\noindent (ii) Since $\, \bar bi\sG b\, $ and
$\, \bar b (i\vec D)^2b\, $
are $SU(3)_{Fl}$ singlet operators and
their expectation values are practically isospin and even
$SU(3)_{Fl}$ invariant, one obtains, as stated before,
$\tau (B_d)\simeq \tau (B^-)\simeq \bar \tau (B_s)$
through order $1/m_b^2$ and likewise
$\tau (\Lambda _b)\simeq \tau (\Xi _b)$. Yet the
meson and baryon lifetimes get differentiated on this level
since $\aver{\mu _G^2}_B > 0 =
\aver{\mu _G^2}_{\Lambda _b,\, \Xi _b}$ and
$\aver{(\vec p_b)^2}_B \neq
\aver{(\vec p_b)^2}_{\Lambda _b}$, see eqs.(20,30).

\noindent (iii) The semileptonic branching ratios of
$\Lambda _b$ and $\Xi _b$ baryons remain unaffected
in order $1/m_b^2$
due to $\aver{\mu _G^2}_{\Lambda _b,\, \Xi _b}\simeq 0$,
whereas for $B$ mesons it is (slightly) reduced \cite{BUV};
i.e., one finds
$$BR_{SL}(B) < BR_{SL}(\Lambda _b) +
{\cal O}(1/m_b^3)\; . $$
Beyond order $1/m_b^2$, however, this relation is changed, as
discussed later on.

\subsection{$B^-$ vs. $B_d$ Lifetimes}
Differences in
$\tau (B^-)$ vs. $\tau (B_d)$ are generated by the local
dimension-six four-quark
operators $(\bar b_L\gamma _{\mu}q_L)
(\bar q_L\gamma _{\nu}b_L)$
which explicitely depend on the
light quark flavours. Such corrections are of order
$1/m_b^3$. Based on this scaling law one can
already infer from the observed $D$ meson lifetime ratios that the
various $B$ lifetimes will differ by no more than 10\% or so.
Assuming factorization for the expectation values of these
four-fermion operators, i.e. applying eq.(22), one obtains
$$\frac{\Gamma (B_d)-\Gamma (B^-)}{\Gamma (B)}
\sim \frac{\Gamma _{nonspect}(B)}{\Gamma _{spect}(B)}
\propto \frac{f_B^2}{m_b^2}\; ,  \eqno(47)$$
as discussed above.

It turns out that WA can change the $B$ lifetimes by
no more than1\% or so; due to interference with the spectator
reaction, they could conceivably
{\em prolong} the $B_d$ lifetime
relative to the $B^-$ lifetime, rather than reduce it. The dominant
effect is clearly provided by PI, which produces a negative
contribution to the $B^-$ width:
$$\Gamma (B^-)\simeq \Gamma _{spect}(B)+
\Delta \Gamma _{PI}(B^-)\eqno(48a)$$
$$ \Delta \Gamma _{PI}(B^-)\simeq \Gamma _0\cdot
24\pi ^2\frac{f_B^2}{m_b^2}
\left[ c_+^2-c_-^2+\frac{1}{N_C}(c_+^2+c_-^2)\right]
\; .\eqno(48b)$$
Eq.(48b) exhibits an intriguing result:
$\Delta \Gamma _{PI}(B^-)$ is positive for $c_+=1=c_-$, i.e.
PI acts {\em constructively} and thus would shorten the
$B^-$ lifetime in the {\em absence} of radiative QCD corrections
\footnote{This shows that the term of `Pauli Interference' should
{\em not} be construed as implying that the interference
is a priori destructive.}.
Including those as evaluated on the leading-log and
next-to-leading-log level --  $c_+\simeq 0.84$, $c_-\simeq 1.42$ --
turns PI into a {\em destructive} interference prolonging
the $B^-$ lifetime, albeit by a tiny amount only at this point.
In eq.(48b) only ultraviolet renormalization has been incorporated.
Hybrid renormalization \cite{HYBRID}
down to the hadronic scale
$\mu _{had}$ amplifies this effect considerably; one
obtains
$$ \Delta \Gamma _{PI}(B^-)\simeq \Gamma _0\cdot
24\pi ^2\frac{f_B^2}{M_B^2}\kappa ^{-4}
\left[ (c_+^2-c_-^2)\kappa ^{9/4}+\frac{c_+^2+c_-^2}{3}
-\frac{1}{9}(\kappa ^{9/2}-1)(c_+^2-c_-^2)\right] \, , $$
$$\kappa \equiv \left[ \frac{\as (\mu _{had}^2)}
{\as (m_b^2)}\right] ^{1/b}, \;
b=11-\frac{2}{3} n_F \eqno(49)$$
Putting everything together one finds:
$$\frac{\tau (B^-)}{\tau (B_d)}\simeq
1+0.05 \cdot \frac{f_B^2}{(200\; MeV)^2}\; , \eqno(50)$$
i.e. the lifetime of a charged $B$ meson is predicted to
definitely {\em exceed} that of a neutral $B$ meson by
typically several percent.

Three comments are in order for properly evaluating
eq.(49):

\noindent $\bullet$ Although WA could conceivably prolong the
$B_d$ lifetime as stated before, its numerical significance pales
by comparison to that of PI. PI acts destructive in
$B^-$ decays once radiative QCD corrections are included to the best
of our knowledge. The lifetime of $B^-$ mesons therefore has to
exceed that of $B_d$ mesons.

\noindent $\bullet$ While the sign of the effect is predicted in an
unequivocal manner, its magnitude is not. The main uncertainty in
the prediction on $\tau (B^-)/\tau (B_d)$ is given by our present
ignorence concerning the size of $f_B$.

\noindent $\bullet$ Even a precise measurement of the lifetime ratio
$\tau (B^-)/\tau (B_d)$ would not automatically result in an
exact determination of $f_B$ by applying eq.(50). For corrections
of order $1/m_b^4$ have not (or only partially -- see the discussion
in the
preceding subsection) been included. Unless those have been
determined, one cannot extract the size of $f_B$ even from
an ideal measurement with better than a roughly 15 \%
uncertainty.

\subsection{$B_s$ Lifetimes}
Very little $SU(3)_{Fl}$ breaking is anticipated between the $B_d$
and $B_s$ expectation values of the two dimension-five operators.
\footnote{This issue will be addressed in some detail in the later
discussion of $D^0$ vs. $D_s$ lifetimes.} Among the contributions
from dimension-six operators
WA affects $B_d$ and $B_s$ lifetimes somewhat differently due to
different colour factors for these two decays. Yet it is quite irrelevant
in either case. Thus one predicts the $B_d$ and the average $B_s$
lifetimes to practically coincide:
$$\tau (B_d) \simeq \bar \tau (B_s) \pm {\cal O}(1\%)
\eqno(51)$$
The term `average $B_s$ lifetime' is used for a reason:
$B_s - \bar B_s$ oscillations generate two neutral beauty mesons
carrying strangeness that differ in their mass as well as in their
lifetime. Due basically to $m_t^2 \gg m_c^2$ one finds
$\Delta m(B_s) \gg \Delta \Gamma (B_s) \neq 0$. While
$\Delta m(B_s)$ can be calculated in terms of the expectation value
of the {\em local} four-fermion operator
$(\bar b_L\gamma _{\mu }q_L)(\bar b_L\gamma _{\mu }q_L)$
(since $m_b \ll m_t$), the situation is more complex for
$\Delta \Gamma (B_s)$, since the underlying operator is nonlocal.
One can however apply a heavy quark expansion; to lowest
nontrivial
order one obtains \cite{BSBS}
$$\frac{\Delta \Gamma (B_s)}{\bar \Gamma (B_s)}
\equiv \frac{\Gamma (B_{s,short})-\Gamma (B_{s,long})}
{\bar \Gamma (B_s)}\simeq 0.18\cdot
\frac{(f_{B_s})^2}{(200 \, MeV)^2} \eqno(52)$$
for $f_{B_s}$ not much larger than 200 MeV. Comparing
eqs.(50-52) leads to the intriguing observation that the largest
lifetime difference among $B^-$, $B_d$ and $B_s$ mesons is
generated by a very subtle source:
$B_s - \bar B_s$ oscillations! There are also two different
lifetimes for neutral $B$ mesons without strangeness; yet $\Delta
\Gamma (B_d)$
is suppressed by $\sim \sin \theta ^2_C$ relative to
$\Delta \Gamma (B_s)$.

One can search for the existence of two different $B_s$ lifetimes by
comparing $\tau (B_s)$ as measured in
$B_s \ra \psi \eta /\psi \phi$ on one hand and in
$B_s\ra l \nu X$ on the other. The
former decay predominantly leads to a CP even final state
and thus would,
to good accuracy,  reveal $\tau (B_{s,short})$; the latter
exhibits the
average lifetime $\bar \tau (B_s)=
[\tau (B_{s,long} - \tau (B_{s,long})]/2$; for semileptonic $B_s$
decays involve the CP even and odd components in a nearly
equal mixture. Thus
$$\tau (B_s \ra l \nu D^{(*)}) -
\tau (B_s\ra \psi \eta /\psi \phi ) \simeq
\frac{1}{2} [\tau (B_{s,long})-\tau (B_{s,short})]\simeq $$
$$\simeq 0.09\cdot
\frac{(f_{B_s})^2}{(200 \, MeV)^2}.  \eqno(53)$$
 Whether an effect of this size is large enough to be ever observed
in a real experiment, is doubtful. Nevertheless one should search
for it even if one has sensitivity only for a 50\% lifetime difference
or so. For while eq.(52) represents the best presently available
estimate, it is not a `gold-plated' prediction. It is conceivable that the
underlying computation {\em underestimates} the actual lifetime
difference!

\subsection{$B_c$ Lifetime}
$B_c$ mesons with their two heavy constituents
-- $B_c=(b\bar c)$ -- represent a highly intriguing system that
merits special efforts to observe it. One expects a
rich spectroscopy probing the inter-quark potential at distances
intermediate to those that determine quarkonia spectroscopy in the
charm and in the beauty system. The Isgur-Wise function for the
striking channel $B_c\ra l \nu \psi$ can be computed. What is most
relevant for our discussion here is that its overall decays, and thus
its lifetime, reflect the interplay of three classes of transitions,
namely
the decay of the $b$ quark, that of the $\bar c$ (anti)quark and WA
of $b$ and $\bar c$:
$$\Gamma (B_c)\simeq \Gamma _{b\ra c} (B_c) +
\Gamma _{\bar c \ra \bar s}(B_c) + \Gamma _{WA}(B_c) \eqno(54)$$
While $b\ra c$ and $\bar c \ra \bar s$ transitions do not
interfere with each other in any practical way and one can thus
cleanly separate their widths, the situation is much more
delicate concerning $\Gamma _{b \ra c}$ and
$\Gamma _{WA}$, as briefly explained later. Yet for the
moment I ignore the latter although
its helicity and wavefunction suppressions represented by the
factors
$m_c^2/m_b^2$ and $f^2_{B_c}/m_b^2$ are relatively mild
($f_{B_c} \simeq 400-600$ MeV \cite{QUIGG})
and partially offset by the
numerical factor $16\pi ^2$ reflecting the enhancement of two-body
phase space -- relevant for WA -- over three-body phase space
appropriate for the spectator decay.

Concerning the other two transitions one would naively expect
the $\bar c$ decay to dominate:
$\Gamma _{\bar c\ra \bar s}(B_c) \sim \Gamma (D^0)
\simeq (4\times 10^{-13}\, sec)^{-1} >
\Gamma _{b \ra c}(B_c) \sim \Gamma (B) \simeq
(1.5\times 10^{-12}\, sec)^{-1}$. It had been suggested that for
a tightly bound system like $B_c$ the decay width should be
expressed not in terms of the usual quark masses (whatever
they are), but instead in terms of effective quark masses
reduced by something like a binding energy
$\mu _{BE} \sim 500$ MeV.
{\em If} so, then the $B_c$ width would be reduced considerably
since due to $\Gamma _Q \propto m_Q^5$ one would find
a relative reduction
$(\Gamma _Q + \Delta \Gamma _Q)/\Gamma _Q \sim
(m_Q - \mu _{BE})^5/m_Q^5 \sim 1 -5\mu _{BE}/m_Q$.
Even more significantly, beauty decays would become more
abundant than charm decays in the $B_c$ transitions since the
binding energy constitutes a higher fraction of $m_c$ than
of $m_b$: $\mu _{BE}/m_c > \mu _{BE}/m_b$.
However this conjecture that might look quite
plausible at first sight, turns out to be fallacious!
For it is manifestly
based on the existence of nonperturbative corrections of
order $1/m_Q$;
yet as discussed in Sect.4 those are absent in fully integrated
decay rates like lifetimes due to a non-trivial cancellation between
$1/m_Q$ contributions from initial-state and final-state radiation!
The leading corrections arise in order $1/m_Q^2$ and
they enter through the first two terms on the
right-hand-side of eq.(54):
$$\Gamma _{b \ra c}(B_c) = \Gamma _{b \ra c,decay}(B_c)
+ {\cal O}(1/m_b^3) \eqno(55a)$$
$$\Gamma _{\bar c \ra \bar s}(B_c) =
\Gamma _{\bar c \ra \bar s,decay}(B_c)
+ {\cal O}(1/m_c^3) \eqno(55b)$$
The quantities $\Gamma _{b \ra c,decay}(B_c)$ and
$\Gamma _{\bar c \ra \bar s}(B_c)$ are defined as in
eqs.(45) and (68); i.e., the differences between
$\Gamma _{decay}(B)$ and $\Gamma _{b \ra c,decay}(B_c)$
enter through the different expectation values of the
same operators $\bar QQ$ and $\bar Q\frac{i}{2}\sG Q$,
and likewise for
$\Gamma _{decay}(D)$ (to be discussed in detail
in Sect.7) vs. $\Gamma _{\bar c \ra \bar s,decay}(B_c)$.
For the $B_c$ meson provides a different environment for
quark decay than either $B$ or $D$ mesons. The relevance of
that difference is illustrated by eqs.(28,29).
Similarly one expects $\aver{(\vec p_b)^2}_{B_c}$,
$\aver{(\vec p_c)^2}_{B_c}$ to differ from
$\aver{(\vec p_b)^2}_B$, $\aver{(\vec p_c)^2}_D$.

While there are large corrections of order $1/m_c^2$ in
$\Gamma _{\bar c\ra \bar s,decay}(B_c)$ that reduce the
corresponding semileptonic branching ratio considerably,
they largely cancel against each other in the total width.
Therefore
$$ \Gamma _{\bar c\ra \bar s,decay}(B_c) \sim
\Gamma _{decay}(D) \simeq \Gamma (D^0) \eqno(56)$$

Writing the $\Delta B=1$ width of the $B_c$ meson as a
simple incoherent sum $\Gamma _{b\ra c}+\Gamma _{WA}$
actually represents an oversimplification. For there
arises considerable interference between higher order
WA and spectator processes. Yet for the purposes of
our discussion here such effects can be ignored;
they will be discussed in detail in a forthcoming
publication. The $1/m_b^2$ contributions to
$\Gamma _{b\ra c,decay}(B_c)$ are small, namely around
a few percent. The numerical impact of
$\Gamma _{WA}(B_c \ra X_{\bar c s})$, which is formally
of order $1/m_b^3$, is nevertheless sizeable due to the
large decay constant, the merely mild helicity suppression,
given by $m_c^2/m_b^2$, and the enhancement
of WA by a factor $\sim 16\pi ^2$ due to its two-body
kinematics.  One finds:
$$\Gamma _{b \ra c}(B_c) \geq \Gamma _{b\ra c}(B)\; .
\eqno(57)$$
Comparing eqs.(56,57) one concludes that
the naive expectation turns out to be basically
correct, i.e. $\tau (B_c)$ is well below $10^{-12}$ sec and charm
decays
dominate over beauty decays with
an ensuing reduction in the `interesting' branching ratios
like $B_c\ra l\nu \psi$ or $B_c\ra \psi \pi$.

\subsection{Beauty Baryon Lifetimes}
It is quite natural to assume that the kinetic energy of the $b$ quark
is practically the same inside the $\Lambda _b$
and $\Xi _b$ baryons:
$$\aver{(\vec p _b)^2}_{\Lambda _b}\simeq
\aver{(\vec p _b)^2}_{\Xi _b} \eqno(58a)$$
Together with eqs.(15,17a) this yields
$$\matel{\Lambda _b}{\bar bb}{\Lambda _b}=
\matel{\Xi _b}{\bar bb}{\Xi _b} + {\cal O}(1/m_b^3)
\eqno(58b)$$
The leading {\em differences} among the $\Lambda _b$ and
$\Xi _b$ lifetimes
then arise in
order $1/m_b^3$: they are generated by four-quark
operators analogous to those
that had already been identified in the phenomenological
studies of charm baryons\cite{BARYONS1,BARYONS2,BARYONS3}.
Some complexities arise, though, due to the
presence of the two transitions $b\ra c \bar ud$
and $b\ra c \bar cs$; one finds
$$\Gamma (\Lambda _b) = \Gamma _{decay}(\Lambda _b) +
\Gamma _{WS}(\Lambda _b)
-|\Delta \Gamma _{PI,-}(\Lambda _b, b\ra c\bar ud)| \eqno(59a)$$
$$\Gamma (\Xi ^0 _b) = \Gamma _{decay}(\Xi _b) +
\Gamma _{WS}(\Xi _b)
- |\Delta \Gamma _{PI,-}(\Xi _b, b\ra c\bar cs)|
\eqno(59b)$$
$$\Gamma (\Xi ^- _b) = \Gamma _{decay}(\Xi _b)
-|\Delta \Gamma _{PI,-}(\Xi _b, b\ra c \bar ud)|
-|\Delta \Gamma _{PI,-}(\Xi _b, b\ra c\bar cs)| \eqno(59c)$$
where \footnote{The channel $b \ra \tau \nu q$ has been ignored
here for simplicity.}
$$\Gamma _{decay}(\Lambda _b/\Xi _b)
\equiv 2 \Gamma _{SL,decay}(\Lambda _b/\Xi _b)+
\Gamma _{NL,decay}(\Lambda _b/\Xi _b) \eqno(60a)$$
$$\frac{\Gamma _{SL,decay}(\Lambda _b/\Xi _b)}{\Gamma _0} =
\matel{\Lambda _b/\Xi _b}{\bar bb}{\Lambda _b/\Xi _b}_{norm}\cdot
\eta _{SL}I_0(x,0,0) \eqno(60b)$$
$$\frac{\Gamma _{NL,decay}(\Lambda _b/\Xi _b)}{\Gamma _0} =
\matel{\Lambda _b/\Xi _b}{\bar bb}{\Lambda _b/\Xi _b}_{norm}\cdot
A_0\cdot [I_0(x,0,0)+ \rho _{c\bar c} I_0(x,x,0)]  \eqno(60c)$$
using the same notation as in eqs.(45);
$$\Gamma _{WS}(\Lambda _b, \Xi _b) \equiv 2\tilde \Gamma _0
c_-^2\matel{\Lambda _b/\Xi _b}{\bar b_L\gamma _{\mu}b_L
\bar u_L \gamma _{\mu}u_L}{\Lambda _b/\Xi _b}_{norm}
\eqno(60d)$$
$$\frac{\Delta \Gamma _{PI,-}(\Lambda _b, b\ra c\bar ud)}
{\tilde \Gamma _0}
\equiv - c_+(2c_- - c_+)
\matel{\Lambda _b}
{\bar b_L\gamma _{\mu}b\bar d_L\gamma _{\mu}d_L +
\frac{2}{3}\bar b\gamma _{\mu}\gamma _5b
\bar d_L\gamma _{\mu}d_L}
{\Lambda _b}_{norm} \eqno(60e)$$
$$\frac{\Delta \Gamma _{PI,-}(\Xi _b, b\ra c\bar cs)}
{\tilde \Gamma _0}\equiv
- c_+(2c_- - c_+)
\matel{\Xi _b}
{\bar b_L\gamma _{\mu}b\bar s_L\gamma _{\mu}s_L +
\frac{2}{3}\bar b\gamma _{\mu}\gamma _5b
\bar s_L\gamma _{\mu}s_L}
{\Xi _b}_{norm} \eqno(60f)$$
$$\tilde \Gamma _0 \equiv 48\pi ^2 \Gamma _0 \eqno(60g)$$
$\Delta \Gamma _{PI,-}(\Xi _b, b\ra c\bar ud)$ is obtained
from eq.(60e) by taking the expectation value between
$\Xi _b$ rather than $\Lambda _b$ states. Those expectation
values vanish for the wrong charge-flavour combination; i.e.,
$\matel{\Xi _b^0}{\bar b_L\gamma _{\mu}b\bar d_L\gamma _{\mu}d_L}
{\Xi _b^0}=0=
\matel{\Lambda _b}{\bar b_L\gamma _{\mu}b\bar s_L\gamma _{\mu}s_L}
{\Lambda _b}$.

$\Gamma _{decay} $ includes the naive
spectator term:
$\Gamma _{decay} = \Gamma _{spect} + 1/m_b^2$ contributions; the
latter are practically identical for $\Lambda _b$ and
$\Xi _b$, but differ for $B$ mesons. $\Gamma _{WS}$ denotes
the contribution due to WS
and $\Delta \Gamma _{PI,-}$ the reduction due to
destructive
interference in the channels $b\ra c\bar ud$ and
$b\ra c \bar cs$, respectively.
\footnote{When $|\Delta \Gamma _{PI,-}(b\ra c \bar ud)|$
is used in eq.(59c), one has,
strictly speaking, to evaluate the expectation value for the
state $\Xi _b$.}

On general grounds one thus obtains an inequality:
$$\tau (\Xi _b^-) > \tau (\Lambda _b)\, , \, \tau (\Xi _b^0)
\eqno(61)$$
Naively one might expect $\Gamma (\Xi _b^0) >
\Gamma (\Lambda _b)$ to hold, since the $b\ra c \bar cs$ part
of the width which is reduced by PI in $\Xi _b^0$ decays
is smaller than the $b\ra c \bar ud$ component which suffers
from PI reduction in $\Lambda _b$ decays. On the other hand since
phase space is more restricted for $b\ra c \bar cs$ than for
$b\ra c \bar ud$, one would likewise expect the degree of
(destructive) interference to be higher for the former than
the latter; it is then quite conceivable that actually
$\Gamma (\Xi _b^0) < \Gamma (\Lambda _b)$ holds. The
{\em sign} of the difference in the $\Lambda _b$ and
$\Xi _b^0$ lifetimes therefore provides us with valuable
information on the strong dynamics.

There are two complementary ways to transform these qualitative
predictions into quantitative ones.

\noindent (i) One evaluates the required expectation values
explicitely within a quark model, as expressed in eq.(17b, 23-25).
Since the model also predicts the baryon masses in terms of its
parameters, one can cross check it with the observed spectroscopy.
This will become increasingly relevant in the future, yet at present
provides little guidance.
The expressions given in eqs.(23-25) have to be augmented
by the radiative QCD corrections:
$$\matel{\Lambda _b}{\bar b\Gamma _{\mu}b
\bar q \Gamma _{\mu} q}{\Lambda _b}_{norm} \sim
\frac{1}{4\aver{\mu ^2_G}_B}(M_{\Sigma _b}-
M_{\Lambda _b})
m^*_q F^2_BM_B\cdot \kappa ^{-4}  \eqno(62a)$$
$$\matel{\Omega _b}{\bar b\Gamma _{\mu}b
\bar q \Gamma _{\mu} q}{\Omega _b}_{norm} \sim
\frac{1}{4\aver{\mu ^2_G}_B}(M_{\Sigma _b}-
M_{\Omega _b})
m^*_q F^2_BM_B\cdot \kappa ^{-4}  \eqno(62b)$$
with $\kappa$ as defined in eq.(49).
It should be noted that here -- unlike in the case of meson
decays -- the sign of the PI contribution is quite robust
under radiative corrections: it is proportional to
$- c_+(2c_- - c_+)$ which is negative already for
$c_+=1=c_-$. Furthermore $\Gamma _{WS}$ is proportional
to $c_-^2$ and thus colour-enhanced, since the baryon
wavefunction is purely antisymmetric in colour space.
This prescription yields lifetime differences
of not more than a few percent in eq.(59) and
$$ \frac{\tau (\Lambda _b)}{\tau (B_d)} \simeq 0.9 - 0.95 \; .
\eqno(63a)$$
(ii) As will be discussed in the next subsection, the
pattern predicted for charm baryon lifetimes agrees
with the observations within the experimental and
theoretical uncertainties. Taking this as prima facie
evidence that the heavy quark expansion through order
$1/m_Q^3$ applies -- at least in a semi-quantitative fashion --
already to charm lifetime ratios, one can extrapolate to the
weight of these pre-asymptotic corrections in beauty
decays using scaling like $1/m_Q^2$ and $1/m_Q^3$
(or $f_M^2/m_Q^2$).
That way one
again finds that the differences in eq.(56) amount to not more
than a few to several percent \cite{STONE2}:
$$ \frac{\tau (\Lambda _b)}{\tau (B_d)} \simeq 0.9 \; .
\eqno(63b)$$

Similarly one estimates $\tau (\Xi _b)$ through a
simple $1/m_Q^3$ scaling behaviour from the charm baryon
lifetimes:
$$\frac{\tau (\Xi _b^-)}{\tau (\Lambda _b)} \sim 1.1 \eqno(64a)$$
$$\frac{|\tau (\Xi _b^0)- \tau (\Lambda _b)|}
{\tau (\Lambda _b)} < 0.1 \eqno(64b)$$

Obviously and not surprisingly there is some numerical fuzziness
in these predictions; yet they seem to be unequivocal in stating
that the $B_d$ lifetime exceeds the $\Lambda _b$ lifetime
and the average beauty baryon lifetime by about 10 percent.
However this prediction appears to be in serious (though not yet
quite conclusive) disagreement with the data. If the predictions
were based exclusively on adopting the quark model results for the
baryonic expectation values, one could abandon them in a
relatively light-hearted way: for it should not come as a shocking
surprise that quark model results for baryonic matrix elements can
be off-target by a substantial amount. Yet we have encountered
a more serious problem here: data seem to contradict also the prediction
based on an extrapolation from the observed lifetime pattern in the
charm family; furthermore -- as discussed next -- the observed
lifetime ratios of charm hadrons can be reproduced, within the
expected uncertainties. This allows only one conclusion: if
$\tau (B_d)$ indeed exceeds $\tau (\Lambda _b)$ by 25 - 30 \%,
then a `theoretical price' has to be paid, namely that

\noindent $\bullet$ the charm mass
represents too low of a scale for allowing to go beyond merely
qualitative
predictions on charm baryon (or even meson) lifetimes,
since it appears that corrections of order $1/m_c^4$ and higher
are still important;

\noindent $\bullet$ that the present agreement between theoretical
expectations and data on charm baryon lifetimes is largely
accidental and most likely would not survive in the face of more
precise measurements!

\noindent At the same time an intriguing puzzle arises:

\noindent $\bullet$ Why are the quark model results for the
relevant expectation values so much off the mark for
beauty baryons? It is the deviation from unity in the lifetime
ratios that is controlled by these matrix elements; finding
a 30 \% difference rather than the expected 10 \% then represents
a 300 \% error!

Some new features emerge in $\Omega _b$ decays: since the
$ss$-diquark system forms a spin-triplet, there are
$1/m_b^2$ contributions to the semileptonic and
nonleptonic $\Omega _b$ widths from the chromomagnetic
operator. In order $1/m_b^3$ there arises a destructive
interference in the $b\ra c \bar cs$ channels. However a
detailed discussion of $\Omega _b$ decays seems
academic at the moment.

\subsection{Semileptonic Branching Ratios of Beauty
Hadrons}

As briefly discussed before, one expects the semileptonic branching
ratios for $B_d$ and $B_s$ mesons to practically coincide,
in particular since semileptonic decays probe $\bar \tau (B_s)$,
the average $B_s$ lifetime.

It has already been pointed out that through order
$1/m_b^2$ the expected value for $BR_{SL}(\Lambda _b)$
exceeds that for $BR_{SL}(B)$ by a few percent.
In order $1/m_b^3$ the nonleptonic widths of both $B_d$ and
$\Lambda _b$ states receive new contributions, with
$\Gamma _{NL}(\Lambda _b)$ getting further enhanced relative
to $\Gamma _{NL}(B)$ as expressed in eq.(63b). Thus one
predicts
$$BR_{SL}(\Lambda _b) < BR_{SL}(B_d) < BR_{SL}(B^-)$$
with the inequalities indicating differences of a few to
several percent. If the present trend in the data persists, i.e.
if the total $\Lambda _b$ width exceeds $\Gamma (B)$ by,
say, 20 -25 \%, one would interprete this discrepancy as
meaning that the nonleptonic -- but not the semileptonic --
width has received a unforeseen enhancement. In that case
one expects $BR_{SL}(\Lambda _b)$ to fall below
$BR_{SL}(B_d)$ by $\sim$ 20 \%.

Putting all these observations together one concludes
that the beauty lifetime averaged over $B_d$, $B^-$,
$B_s$ and $\Lambda _b$ states should yield a value that
is {\em smaller} -- by a few percent -- than
$\frac{1}{2} \left[ BR_ {SL}(B_d) + BR_ {SL}(B^-)\right]
\equiv \aver{BR_{SL}(B)}$:
$$\aver{BR_{SL}(b)}< \aver{BR_{SL}(B)}$$

\section{Predictions on Charm Lifetimes}

Considering the wealth of rather precise experimental information
available on the lifetimes of charm hadrons one feels the urge to
apply heavy quark expansions to charm decays as well. Yet in doing
so one has to keep in mind that such a treatment might fail here
for two basic reasons, one of which has just been stated:

\noindent (i) The charm quark mass does not provide a
sufficiently large scale to make the $1/m_c$ expansion
converge quickly. To obtain an estimate of the size of the
expansion parameter $\mu _{had}/m_c$ one can take the
square root of the expression in eq.(35)
representing
the $1/m_c^2$ corrections:
$$\frac{\mu _{had}}{m_c}\sim \sqrt{\frac{\aver {\mu _G^2}_D}
{m_c^2}}\simeq 0.45 \, .\eqno(65)$$
This is not a small number although it is at least smaller
than unity.

\noindent (ii) As
discussed in Sect.4 one has to go beyond
{\em global} duality and invoke {\em local} duality to predict the
decay widths of real hadrons from heavy quark expansions properly
defined in Euclidean space. Yet a priori it is not clear at all
whether contributions from `distant cuts' can be ignored since the
charm quark mass does not exceed ordinary hadronic scales by a
large amount. This concern is a posteriori strengthened by the
following observation: Equating the observed semileptonic width of
$D$ mesons with its theoretical expression through order $1/m_c^2$
(and assuming $|V(cs)| \simeq 1$) leads to the requirement
$"m_c" \simeq 1.6$ GeV. This is however a high value relative
to what is derived from charmonium spectroscopy, namely
$m_c \leq 1.4$ GeV. A difference of $0.2$ GeV in $m_c$ might
appear quite innocuous -- till one realizes that the corresponding
semileptonic width depending on $m_c^5$ changes by a factor of
two when $m_c$ is shifted by those $0.2$ GeV! At this point one
might suspect that corrections of higher order in $1/m_c$ contribute
{\em constructively} boosting the theoretical value. The analysis of
ref.\cite{DIKEMAN} finds however that terms of order $1/m_c^3$
show a
tendency to  {\em decrease} $\Gamma _{SL}(D)$, and their inclusion
does
not help at all in reproducing the measured value of
$\Gamma _{SL}(D)$  with $m_c\simeq 1.4$ GeV. This
suggests that contributions from `distant cuts' which cannot be seen
in any finite order of the $1/m_c$ expansion create this problem.
For a different opinion on this point, see ref.\cite{CHERNYAK}.

I draw the following somewhat tentative conclusions from these
observations:

\noindent $\bullet$ Predictions for the absolute size of charm hadron
lifetimes cannot be trusted.

\noindent $\bullet$ However it is quite conceivable that lifetime
{\em ratios} do not suffer from such a fundamental uncertainty due
to not-so-distant cuts. I will then explore the working hypothesis
that the
ratio of lifetimes can be trusted in principle -- though in
practise only with a quite considerable grain of salt!

The quark level Lagrangian is again well known. On the
Cabibbo-allowed level there is now a single
non-leptonic transition described by
$${\cal L}_W^{\Delta C=1}(\mu = m_c)=
\frac{4G_F}{\sqrt{2}} V_{cs}V_{ud}^*
\{ c_1(\bar s_L\gamma _{\mu}c_L)(\bar u_L\gamma _{\mu}d_L)
+c_2(\bar u_L\gamma _{\mu}c_L)(\bar s_L\gamma _{\mu}d_L)
\}    \eqno(66)$$
where the short-distance coefficients $c_{1,2}$ are now evaluated
at a lowerJscale than in beauty decays yielding
$$ c_1(LL+NLL)\simeq 1.32 \, , \; \;
c_2(LL+NLL)\simeq - 0.58 \eqno(67)$$.

The expressions for the semileptonic and non-leptonic
decay widths of charm hadrons $H_c$
are quite analogous to the ones for beauty hadrons
(though simpler since there is only one non-leptonic
decay class rather than two); through order $1/m_c^2$
they are given by
$$\frac{\Gamma _{SL,decay}(H_c)}{\Gamma _0}=
\matel{H_c}{\bar cc}{H_c}_{norm}\cdot
\left[ I_0(x,0,0) +
\frac{\aver{\mu _G^2}_{H_c}}{m_c^2}
(x\frac{d}{dx}-2)I_0(x,0,0)\right] \; , \eqno(68a)$$
$$\frac{\Gamma _{NL,decay}(H_c)}{\Gamma _0} = N_C\cdot
\matel{H_c}{\bar cc}{H_c}_{norm}\cdot
\left[A_0[I_0(x,0,0) +
\frac{\aver{\mu _G^2}_{H_c}}{m_c^2}
(x\frac{d}{dx}-2)I_0(x,0,0)] - \right.$$
$$\left. 8A_2 \frac{\aver{\mu _G^2}_{H_c}}{m_c^2}
\cdot
I_2(x,0,0,)\right]\; . \eqno(68b)$$
where now
$$\Gamma _0 \equiv
\frac{G_F^2m_c^5}{192 \pi ^3}|V(cs)|^2 \; , \; \;
x=\frac{m_s^2}{m_c^2}\eqno(68c)$$
and the radiative corrections lumped into $A_0$
and $A_2$ are given by the appropriate values for
$c_+$ and $c_-$, see eq. (67). With $x\sim 0.012$ one finds:
$$I_0(x,0,0)|_{x=0.012}\simeq 0.91 \; \; , \; \;
I_2(x,0,0)|_{x=0.012}\simeq 0.96 \; , $$
i.e. there is much less phase space suppression than for
$b\ra c$ transitions.

\subsection{$D^+$ vs. $D^0$ Lifetimes}
Analogously to the case of $\tau (B^-)$ vs. $\tau (B_d)$ the
$D^+$ and $D^0$ lifetimes get first differentiated in
order $1/m_c^3$
when PI and WA intervene. Both affect the lifetime
ratio in the same direction, namely they enhance
$\tau (D^+)$ over $\tau (D^0)$ with the
{\em destructive interference} due
to PI being the {\em dominant} effect. Quantitatively one finds through
order $1/m_c^3$ by employing eqs.(48,49) with the appropriate
substitutions and using $f_D \sim 200$ MeV:
$$\frac{\tau (D^+)}{\tau (D^0)} \sim 2 \eqno(69)$$
For proper perspective one has to keep four observations in mind:

\noindent (i) While the expectation expressed in eq.(69) does not
coincide numerically with the measured value --
$\tau (D^+) /\tau (D^0) = 2.547 \pm 0.043$ -- it agrees with it to
within $\sim 25$ \%. Such a deviation could be ascribed to
$1/m_c^4$ contributions ignored in eq.(69).

\noindent (ii) On the other hand, as discussed before, there is reason
to doubt the reliability and thus validity of heavy quark expansions
for treating nonperturbative corrections in
{\em charm} decays. Yet I
adopt, as already stated, the working hypothesis that a heavy quark
expansion can be employed for treating the {\em ratios} of lifetimes,
though not the lifetimes themselves. Such a conjecture is tested
a posteriori by analysing the whole pattern of the expected and the
observed lifetimes of the various charm hadrons.

\noindent (iii) Even the perturbative corrections
contain sizeable numerical uncertainties. To cite but one glaring
example: the size and the nature of the socalled `hybrid'
renormalization reflecting dynamics between the scales
$m_c$ and $\mu _{had} \sim 0.5 - 1$ GeV is quite important
quantitatively (they considerably enhance the destructive
interference in $D^+$ decays.) Yet a leading-log treatment of those
corrections seems woefully inadequate for dealing with such a small
slice in momentum space.

\noindent (iv) One should note also the following: while WA plays
only a relatively
minor role in inclusive rates (generating only a 10 \% or so
difference in $D^+$ vs. $D^0$ lifetimes), it is likely to play a
considerably more significant role in {\em exclusive} modes.

\subsection{$D_s^+$ vs. $D^0$ Lifetimes}
Since the impact of WA is reduced relative to that of PI in meson
decays, it is natural to compare $\tau (D_s)$ to
$\tau (D^0)$ rather than to $\tau (D^+)$. Such a comparison
-- and of their
semileptonic branching ratios -- touches upon several intriguing
dynamical issues. A priori $\tau (D_s)$ and $\tau (D^0)$ could
differ substantially from each other due to $SU(3)_{Fl}$
breaking and in particular due to a different weight of WA
in the two cases. Yet the heavy quark expansion
strongly suggests
those two lifetimes to be close to each other.

The analysis proceeds in several steps \cite{DS}.
The two operators
contributing in order $1/m_c^2$ are singlets under the
light quark flavours; yet $SU(3)_{Fl}$ breaking could enter
through their expectation values. For the chromomagnetic
operator one has:
$$\frac{1}{M_D}\matel{D^0}{\bar c i\sG c}{D^0}
\simeq \frac{3}{2} (M^2_{D^{*0}}-M^2_{D^0})\eqno(70)$$
$$\frac{1}{M_{D_s}}\matel{D_s}{\bar c i\sG c}{D_s}
\simeq \frac{3}{2} (M^2_{D_s^*}-M^2_{D_s})\eqno(71)$$
Since the measured values for the $D-D^*$ and the $D_s-D_s^*$
mass splittings are practically identical
( $M_{D^{*0}} - M_{D^0}=142.12 \pm 0.07$ MeV and
$M_{D_s^*} - M_{D_s}=141.6 \pm 1.8$ MeV ), the chromomagnetic
operator cannot induce an appreciable difference between
$\tau (D^0)$ and $\tau (D_s)$.
\footnote{The observation that the hyperfine splitting is largely
independant of the flavour of the spectator (anti)quark can be
understood intuitively in quark models.}

Through order $1/m_c^3$ there are four distinct
sources for a difference in $\Gamma (D_s)$ vs.
$\Gamma (D^0)$ exceeding the 1\% level:
(a) The decay $D_s\ra \tau \nu$; (b) PI in those Cabibbo suppressed
$D_s$ decays that are driven by the quark level transition
$c\ra s\bar su$; (c) the effects of $SU(3)_{Fl}$ breaking on the
expectation values of the kinetic energy operator;
(d) WA in nonleptonic $D^0$ and in nonleptonic as well as
semileptonic $D_s$ decays. Corrections listed under
(a), (b) and (d) are generated by dim-6 operators
whereas the much less familiar correction (c) is derived from a
dim-5 operator.

\noindent (a) The width for the decay $D_s\ra \tau \nu$ is
completely determined in terms of the axial decay constant for the
$D_s$ meson:
$$\Gamma (D_s) = \frac{G_F^2m^2_{\tau}f^2_{D_s}M_{D_s}}
{8\pi}|V(cs)|^2(1-m^2_{\tau}/M^2_{D_s})^2 .  \eqno(72)$$
For $f_{D_s}\simeq 210$ MeV one gets
$$\Gamma (D_s\ra \tau \nu ) \simeq 0.03 \Gamma (D^0) .
\eqno(73)$$
This effect necessarily {\em reduces} $\tau (D_s)$ relative to
$\tau (D^0)$.

\noindent (b) In $D_s$ decays PI appears in the
Cabibbo-suppressed
$c\ra s \bar su$ channel. Its weight is expressed in terms of the
matrix elements of the two four-fermion operators
$$\matel{D_s}{(\bar c_L\gamma _{\mu}s_L)
(\bar s_L\gamma _{\mu}c_L)}{D_s}, \; \;
\matel{D_s}{(\bar c_L\gamma _{\mu}\lambda ^as_L)
(\bar s_L\gamma _{\mu}\lambda ^ac_L)}{D_s} \eqno(74)$$
with known coefficients that are computed
perturbatively (including the `hybrid' renormalization of these
operators
down from the scale $m_c$). The effect of PI is
most reliably estimated from the observed difference in the
$D^+$ and $D^0$ widths. It is easy to see that the structure of
the operators responsible for PI in $D_s$ decays is exactly the
same as in $D^+$ decays if one replaces the $d$ quark with the
$s$ quark and adds the extra factor $\tan ^2\theta _C$; it is then
{\em destructive} as well. Thus one arrives at:
$$\delta \Gamma _{PI}(D_s) \simeq S\cdot \tan ^2\theta _C
(\Gamma (D^+)-\Gamma (D^0)) \simeq -S\cdot 0.03
\Gamma (D^0)\eqno(75)$$
where the factor $S$ has been introduced to allow for $SU(3)_{Fl}$
breaking in the expectation values of the four-fermion operators.
The factor $S$ is expected to exceed unity somewhat; in an
factorization ansatz it is given by the ratio
$f^2_{D_s}/f^2_D \simeq 1.4$, see eq.(39). Thus
$$ \delta \Gamma _{PI}(D_s) \simeq - 0.04 \Gamma (D^0) \, .
\eqno(76)$$

\noindent (c) The impact of the
{\em chromomagnetic}
operator on the $D_s-D^0$ lifetime ratio can be derived
from the hyperfine splitting. Inserting the observed meson
masses one obtains
$$ \matel{D^0}{\bar c \frac{i}{2} \sG c}{D^0}_{norm}
\simeq 0.413 \pm 0.002 \eqno(77a)$$
 $$ \matel{D_s}{\bar c \frac{i}{2} \sG c}{D_s}_{norm}
\simeq 0.433 \pm 0.006 \eqno(77b)$$
from which one can conclude: there is -- not surprisingly --
very little $SU(3)_{Fl}$ breaking in these matrix elements and
a sizeable difference in $\tau (D_s)$ vs. $\tau (D^0)$
{\em cannot} arise from this source.

\noindent The second dim-5 operator generating
$1/m_c^2$ corrections is the kinetic energy operator
$\bar c (i\vec D)^2c$ where $\vec D$ denotes the covariant
derivative. Its expectation value describes the largely
non-relativistic
motion of the charm quark in the gluon background field
inside the charm
hadron. It reflects Lorentz
time dilation and thus prolongs the {\em hadron} lifetime. On
general grounds one expects it to extend $\tau (D_s)$ over
$\tau (D^0)$, as seen as follows. The spatial wavefunction should
be more concentrated around the origin for $D_s$ than for
$D^0$ mesons. This implies, via the uncertainty principle, the
mean value of $(\vec p_c)^2$ to be larger for $D_s$ than for
$D^0$; in other words the charm quark undergoes more Fermi
motion as constituent of $D_s$ than of $D$ mesons. The lifetime of
the charm quark is then prolonged by time dilation to a higher
degree inside $D_s$ than inside $D^0$ mesons.

\noindent While the trend of this effect is quite transparent, its
size is not yet clear. The relevant matrix elements
$\matel{D_s}{\bar c (i\vec D)^2 c}{D_s}$ and
$\matel{D}{\bar c (i\vec D)^2 c}{D}$ will be determined from
QCD sum rules and lattice simulations in the future. Yet in the
meantime one can estimate their size from the meson
masses in the charm and beauty family according to the following
prescription:
$$ \matel{D_s}{\bar c (i\vec D)^2 c}{D_s} -
\matel{D}{\bar c (i\vec D)^2 c}{D} \simeq
\frac{2m_bm_c}{m_b-m_c}
\{ [\aver{M_{D_s}}-\aver{M_D}]-
[\aver{M_{B_s}}-\aver{M_B}]\}
\eqno(78)$$
where as before $\aver{M_{D,D_s,B,B_s}}$ denote the
spin averaged masses. Accordingly one finds
$$\frac{\Delta \Gamma (D_s/D)}{\bar \Gamma}\equiv
\frac{\Gamma _{Fermi}(D_s)- \Gamma _{Fermi}(D)}
{\bar \Gamma } \simeq $$
$$ \simeq -\frac{m_b}{m_c(m_b-m_c)}
\times \{ [\aver{M_{D_s}}-\aver{M_D}]-
[\aver{M_{B_s}}-\aver{M_B}]\} \, .\eqno(79)$$
Since the $B_s^*$ mass has not been measured yet, one
cannot give a specific numerical prediction and has
to content oneself with semi-quantitative statements.
A 10 MeV shift in any of the terms $\aver{M}$
corresponds numerically to the kinetic energy operator
generating approximately a 1\% change in the ratio
$\tau (D_s)/\tau (D^+)$. Invoking our present understanding
of the heavy quark kinetic energy and its relationship to
the hyperfine splitting one arrives at the following
{\em conjecture}:
$$ \frac{\Delta \Gamma (D_s/D)|_{(a)+(b)+(c)}}{\bar \Gamma}
\sim {\cal O}(+\, few\, \% )\eqno(80)$$

\noindent (d) The mechanisms listed above in
(a) - (c) taken together can be expected to extend the
$D_s$ over the $D^0$ lifetime by at most a few percent.
Comparing that estimate with the data allows us to
conclude that WA cannot change $\tau (D_s)/\tau (D^0)$
by more than 10 \% . It is however quite unclear
how to refine this estimate at present. For the quantitative
impact of WA on charm
meson lifetimes in general and on the ratio
$\tau (D_s)/\tau (D^0)$ in particular
is the most obscure theoretical element in the
analysis. The uncertainty centers
mainly on the question of how much the WA amplitude suffers
from helicity suppression.

\noindent In the valence quark approximation
the answer is easily given to {\em lowest} order
in the strong coupling: the WA
amplitude is (helicity) suppressed by the ratio
$m_q/m_c$, where $m_q$ denotes the largest quark mass
in the final state. For a proper QCD treatment one has to use
the {\em current} rather than the larger
{\em constituent} mass, at least for
the $1/m_c^3$ corrections. The WA amplitude is then small in
$D^0$ decays where the helicity factor reads
$m_s/m_c \sim 0.1$ and a fortiori in $D_s$ decays where
only non-strange quarks are present in the final state.
The emission of semi-hard gluons that is included by
summing the leading log terms in the perturbative expansion
cannot circumvent this suppression \cite{MIRAGE}. For such gluon
corrections -- when properly accounted for -- drive the hybrid
renormalization of the corresponding four-fermion operators;
however they preserve the Lorentz structure of the lowest order
term and therefore do not eliminate the helicity suppression.
A helicity allowed amplitude arises in perturbation theory only at
the subleading level of order $\as (m_c^2)/\pi$ and is thus expected
to be numerically insignificant.

\noindent On the other hand {\em nonperturbative} dynamics
can quite naturally vitiate helicity suppression and thus provide
the dominant source of WA. These nonperturbative effects enter
through {\em non-factorizable} contributions to the hadronic
matrix elements, as analyzed in considerable detail in
refs.\cite{WA,DS}. As such we do not know (yet), how to predict
their weight from first principles. However, as shown in
ref.\cite{WA}, a detailed experimental study of the width of
semileptonic decays and their lepton spectra -- in particular
in the endpoint region -- in $D^0$ vs. $D_s$ and/or in $B^0$
vs. $B^+$ decays would allow us to extract the size of the
matrix elements that control the weight of WA in all
inclusive $B$ and $D$ decays. Before such data become
available, we can draw only qualitative, or at best
semi-quantitative conclusions: WA is not expected to
affect the total lifetimes of $D_s$ and $D^0$ mesons by more
than 10 - 20 \%, and their ratio by less.
Furthermore WA does not
necessarily decrease $\tau (D_s)/\tau (D^0)$; due to its
reduced amplitude its leading impact on the lifetime might
be due to its interference with the spectator amplitude and thus
it might even {\em enhance} $\tau (D_s)/\tau (D^0)$!

To summarize our findings on the $D_s-D^0$ lifetime ratio:
(i) $SU(3)_{Fl}$ breaking in the expectation values of the
dim-5 operators generating the leading nonperturbative
corrections of order $1/m_c^2$ can -- due to
`time dilatation' -- increase $\tau (D_s)$ by 3-5 \% over
$\tau (D^0)$. (ii) On the $1/m_c^3$ level there arise
three additional effects. Destructive interference in
Cabibbo suppressed $D_s$  decays increases $\tau (D_s)$
again by 3-5 \%, whereas the single mode
$D_s\ra \tau \nu$ decreases it by 3 \%. The three
phenomena listed so far combine to yield
$\tau (D_s)/\tau (D^0) \simeq 1.0 - 1.07$.
(iii) Any difference over and above that has to be
attributed to WA. Therefore one has to interprete the
measured $D_s-D^0$ lifetime ratio as more or less direct
evidence for WA to contribute no more than 10-20 \% of
the lifetime ratio between charm mesons. This is consistent
with the indirect evidence discussed above that WA does not
constitute the major effect there. (iv) These predictions can
be refined in the future by two classes of more accurate data:
analyzing the {\em difference} in the
semileptonic spectra of charged and neutral mesons in the charm
and beauty family allows to extract the size of the matrix elements
that control the weight of WA; measuring the masses of
$\Lambda _b$ and $B_s^*$ to better than 10 MeV allows to
determine the expectation values of the kinetic energy
operator.

\subsection{Charm Baryon Lifetimes}
The lifetimes of the weakly decaying charm baryons reflect a
complex interplay of different dynamical mechanisms
increasing or decreasing transition rates
\cite{MARBELLA,BARYONS3}:
$$\Gamma (\Lambda _c^+)= \Gamma _{decay}(\Lambda _c^+) +
\Gamma _{WS}(\Lambda _c^+) -
|\Delta \Gamma _{PI,-}(\Lambda _c)|   \eqno(81a)$$
$$\Gamma (\Xi _c^0)= \Gamma _{decay}(\Xi _c^0) +
\Gamma _{WS}(\Xi _c^0) +
|\Delta \Gamma _{PI,+}(\Xi _c^0)|   \eqno(81b)$$
$$\Gamma (\Xi _c^+)= \Gamma _{decay}(\Xi _c^+) +
|\Delta \Gamma _{PI,+}(\Xi _c^+)| -
|\Delta \Gamma _{PI,-}(\Xi _c^+)|   \eqno(81c)$$
The explicit expressions for $\Gamma _{decay}$, $\Gamma _{WS}$
and $\Delta \Gamma _{PI,-}$ are obtained from eqs.(60) by the
obvious substitutions; for $\Delta \Gamma _{PI,+}$ one finds
$$\Delta \Gamma _{PI,+}(\Xi _c)\equiv
\frac{G_F^2|V(cs)|^2m_c^2}{4\pi}
c_+(2c_- + c_+)
\matel{\Xi _c}
{\bar c_L\gamma _{\mu}c\bar s_L\gamma _{\mu}s_L +
\frac{2}{3}\bar c\gamma _{\mu}\gamma _5 c
\bar s_L\gamma _{\mu}s_L}
{\Xi _c}_{norm} \eqno(82)$$
 On rather general grounds one then concludes:
$$\tau (\Xi _c^0) < \tau (\Xi _c^+)\; , \; \; \;
\tau (\Xi _c^0) < \tau (\Lambda _c^+) \eqno(83)$$
To go further requires calculating the relative weight of
matrix elements of the four-quark operators.
At present this can be done only by
using quark wavefunctions as obtained from a potential ansatz
(or from the quark model). In doing so one also has to
pay proper attention to the normalization point appropriate
for such an evaluation, i.e., one has to include ultraviolet as
well as hybrid renormalization. One then arrives at the
numbers quoted in Table 2.
It has been argued \cite{MARBELLA}
that in these expressions one should not
use the `real' value for the decay constant, as expressed by
$f_D$, and the
observed mass splitting, but solely the leading contribution
in a $1/m_Q^3$ expansion: $F_D \sim 400$ MeV and
$M_{\Sigma ^+_c} - M_{\Lambda _c}\sim 400$ MeV.
One reason for that is self-consistency since the widths have
been calculated through order $1/m_Q^3$ only; the other --
and maybe the more telling one -- is based on the needs of
phenomenology: for otherwise one cannot reproduce the
observed {\em magnitude} of the lifetime differences.
The predictions
given in Table 2 agree remarkably well with the data
within the experimental and theoretical uncertainties.
\footnote{One should point out that those predictions
were made before these data became available; data at that
time suggested considerably larger lifetime ratios.}
In the comparison of the $\Lambda _c$ and $D$ lifetimes
one should note the following: WS and PI {\em counteract}
each other in $\Gamma (\Lambda _c)$, see eq.(81a);
nevertheless $\Gamma (\Lambda _c)\sim 2\cdot \Gamma (D)$
can be obtained. This results from three effects:
(i) WA is very much reduced in $D$ decays. (ii) The
{\em baryonic} expectation values of the four-quark
operators are evaluated, as stated above, with the
`static' decay constant, which is considerably larger
than $f_D$ used for the {\em mesonic} expectation values.
(iii) The $1/m_c^2$ contributions enhance
$\Gamma (\Lambda _c)$ over $\Gamma (D)$, as briefly discussed
below.

Unfortunately there appears now a rather unpleasant
`fly in the ointment': as discussed above, extrapolating from
these prima facie successful predictions to the lifetimes of
beauty baryons leads to a less-than-successful prediction on
$\tau (\Lambda _b)/\tau (B_d)$.

The decays of $\Omega _c$ baryons require -- and deserve --
a separate treatment: the $ss$ diquarks carry spin one
and the resulting spin-spin interactions of $c$ with $ss$
lead to new effects. To be more specific: through order
$1/m_c^3$ one has
$$\Gamma (\Omega _c) = \Gamma _{decay}(\Omega _c) +
|\Delta \Gamma _{PI,+}(\Omega _c)| \eqno(84)$$
with both quantities on the right-hand-side of eq.(84)
differing from the corresponding ones for $\Lambda _c$
or $\Xi _c$ decays.
(i) Firstly,
$\matel{\Omega _c}{\bar c i\sG c}{\Omega _c}\neq 0
= \matel{\Lambda _c}{\bar c i\sG c}{\Lambda _c}$, see
eqs.(17b,17c,35). Secondly, it is quite conceivable that
these spin-spin interactions can create an
appreciable difference in the
kinetic energy of the $c$ quark inside $\Omega _c$ and
$\Lambda _c$ baryons.  Thus there arise differences of order
$1/m_c^2$ in the total as well as in the
semileptonic widths of $\Omega _c$ and $\Lambda _c$ baryons:
$$\Gamma _{decay}(\Omega _c) \neq \Gamma _{decay}(\Lambda _c)
\; \; , \; \;
\Gamma _{SL,decay}(\Omega _c) \neq \Gamma _{SL,decay}(\Lambda _c)
\eqno(85)$$
(ii) These spin-spin interactions also
affect the expectation
values of the dimension-six four-fermion operators that control
the strength of WS and the interference effects, see eqs.(24,25),
in order $1/m_c^3$.

\noindent Taking everything together one estimates the
$\Omega _c$ to be the shortest lived charm baryon, as
shown in Table 2. This is quite remarkable since it means that
the constructive interference in $\Omega _c$ decays outweighs
the {\em combined} effect of constructive interference and
WS in $\Xi _c^0$ decays. It is actually not unexpected on
intuitive grounds: in $\Omega _c$ transitions interference
can happen with {\em both} constituent $s$ quarks whereas in
$\Xi _c$ baryons there is only one such $s$ quark.
This shows that inclusive weak decay rates
can be harnessed to probe the internal structure
of charm baryons in a novel way.

\subsection{Semileptonic Branching Ratios of Charm
Hadrons}
As expected, the ratio of the semileptonic $D^+$ and $D^0$
branching ratios agrees, within the errors, with the ratio
of their lifetimes:
$$2.23 \pm 0.42=
\frac{BR_{SL}(D^+)}{BR_{SL}(D^0)}\simeq
\frac{\tau (D^+)}{\tau (D^0)}=\, 2.547\, \pm \, 0.043
\eqno(86)$$
Also the {\em absolute} size of, say, $BR_{SL}(D^0)$ is of
interest. If the $D^+ - D^0$ lifetime difference is
generated mainly by a destructive interference
affecting $D^+$ transitions, then $D^0$ decays should
proceed in a largely normal way and the
semileptonic branching ratio of $D^0$ mesons should be at its
`normal' value. In a parton model description one finds
$BR(D\ra l\nu X_s) \equiv BR(c\ra l\nu s) \sim
16\%$. If that number represented the proper
yardstick for normal decays, one would actually conclude
that $D^+$ decays more or less normally, whereas the
nonleptonic decays of $D^0$ are strongly enhanced; i.e.,
it would imply that it is actually WA that provides the
dominant mechanism for the lifetime difference. This
long-standing `fly in the ointment' can be removed now.
For at order $1/m_Q^2$ the chromomagnetic operator
$\bar Qi\sG Q$ generates an  isoscalar enhancement in the
nonleptonic widths of the charged and neutral mesons
\cite{BUV}.
It turns out that the corresponding reduction in the
semileptonic branching ratios of $B$ mesons is rather
insignificant, namely $\sim$ 2 \% . Yet for charm mesons
this reduction is quite large, namely around
40 \%, since it is
scaled up by $(m_b/m_c)^2$ and less reduced
by colour factors. The semileptonic branching
ratio for $D$ mesons is thus around 9 \% through
order $1/m_c^2$ -- before the dimension-six
four-fermion operators reduce
$\Gamma _{NL}(D^+)$ by a factor of almost two and
enhance $\Gamma _{NL}(D^0)$ by a moderate amount.
As discussed above, we have to take these numbers for
charm transitions with a grain of salt; on the other hand
the emerging pattern in the ratio of semileptonic branching
ratios as well as their absolute size is self-consistent!

It should be apparent from the discussion on $\tau (D_s)$
vs. $\tau (D^0)$ that the semileptonic widths and branching
ratios of $D_s$ and $D^0$ mesons practically coincide
through order $1/m_c^2$. In order $1/m_c^3$ WA
can contribute to $\Gamma _{SL}(D_s)$ (as well as to
$\Gamma _{NL}(D_s)$ and $\Gamma _{NL}(D^0)$).
Yet from $\tau (D_s)\simeq \tau (D^0)$ one infers that the
relative weight of WA is quite reduced; therefore one expects
$BR_{SL}(D_s)$ to differ from $BR_{SL}(D^0)$ by not more
than 10\% or so in {\em either} direction:
$$1\, \pm \sim {\rm few}\% = \frac{\tau (D_s)}{\tau (D^0)}
\simeq \frac{BR_{SL}(D_s)}{BR_{SL}(D^0)}=1\, \pm \sim 10\%
\eqno(87)$$

The semileptonic width is not universal for all
charm hadrons due to ${\cal O}(1/m_c^2)$
contributions even when one ignores Cabibbo suppressed
channels. One actually estimates
$$ \Gamma _{SL}(D) / \Gamma _{SL}(\Lambda _c) /
\Gamma _{SL}(\Omega _c) \, \sim \,
1/1.5/1.2 \, .\eqno(88)$$
Therefore the ratio of
semileptonic branching ratios of mesons and
baryons does not faithfully reproduce the ratio of their
lifetimes. Instead one expects
$$BR_{SL}(\Lambda _c) > BR_{SL}(D^0)\times
\tau (\Lambda _c)/\tau (D^0)
\simeq 0.5\, BR_{SL}(D^0) \, . \eqno(89)$$

\section{What Have We Learned and What Will We Learn?}

Second generation theoretical technologies have been developed for
treating heavy-flavour decays that are directly related to QCD
without the need for invoking a `deus ex machina'. As far as
{\em inclusive} heavy-flavour decays are concerned, the relevant
technology is based on an operator expansion in inverse powers
of the heavy quark mass \footnote{To repeat it one last time: This
is {\em not} HQET!}. It allows to express the leading nonperturbative
corrections through the expectation values of a small number of
dimension-five and -six operators. Basically all such matrix
elements relevant for {\em meson} decays can reliably be related to
other
observables; this allows to extract their size in a
model-independant way.
Baryons, however, possess a more complex internal
structure, which becomes relevant for their decays
in order $1/m_Q^3$. At present one has
to rely on quark model calculations to determine
the expectation values of the dimension-six operators relevant for
lifetime differences. The numerical results of such computations
are of dubious reliability; accordingly predictions for lifetime ratios
involving heavy-flavour {\em baryons} suffer from larger
uncertainties than those involving only mesons. There is
also a positive side to this, though: measuring the lifetimes (and if
at all possible also the semileptonic branching ratios) of the
various heavy-flavour baryons sheds new light onto their
internal strong dynamics. Comprehensive studies of inclusive
weak decays in the charm baryon family are particularly
promising from a practical point
for such studies on hadronic structure; for the
pre-asymptotic corrections are much larger here than
for beauty baryons.

At present there exists one rather glaring phenomenological
problem for the heavy quark expansion and -- not surprisingly, as
just indicated -- it concerns baryon decays: the observed
$\Lambda _b$ lifetime is clearly
shorter than predicted relative to the $B_d$ lifetime.
Unless future measurements move it up significantly, one
has to pay a theoretical price for that failure. As stated above,
for the time being one
has to employ a quark model to determine the size of the
baryonic expectation values of the relevelant dimenion-six
operators. It is not surprising per se that quark model computations
can yield numerically incorrect results, in particular  when sizeable
cancellations occur between different contributions. The more
serious problem is provided by the following aspects: the
analoguous treatment of charm baryons, using the same quark
models, had yielded predictions
that give not only the correct pattern, but also numerical values for
the lifetime ratios which are surprisingly close to the data
considering the uncertainties due to higher order terms in the
$1/m_c$ expansion. Furthermore scaling these charm lifetime
ratios up to the beauty scale, using a $1/m_Q^2$ and $1/m_Q^3$
behaviour for the pre-asymptotic corrections from the
dimension-five and -six operators, respectively, also tells us
that the $\Lambda _b$ lifetime should be shorter than the
$B_d$ lifetime by no more than 10 \%.
To the degree that the observed value for
$\tau (\Lambda _b)/\tau (B_d)$ falls below 0.9 one has to draw the
following conclusion: one cannot trust
the numerical results of quark model
calculations for {\em baryonic} matrix elements -- not very
surprising by itself; yet furthermore and more seriously
it would mean that $1/m_Q^4$ or even higher order
contributions are still relevant in charm baryon decays before
fading away for beauty decays.
Then one had
to view the apparently successful predictions on the lifetimes
of charm baryons as largely coincidental -- a quite sobering
result!

A {\em future} discrepancy between the predictions on
$\tau (B^+)/\tau (B_d)$ or
$\tau (B_d)/\bar \tau (B_s)$ and the data \footnote{It has to be
stressed again that $\bar \tau (B_s)$ refers to the algebraic
average of the two $B_s$ lifetimes.} -- in particular an
observation that the lifetime for $B^+$ mesons is definitely
{\em shorter}
than for $B_d$ mesons -- would have quite fundamental
consequences. For the leading deviation of these ratios from
unity arises at order $1/m_b^3$ and should provide a good
approximation since the expansion parameter is small:
$\mu _{had}/m_b \sim 0.1$. The size of this term is given by
the expectation value of a four-fermion operator expressed in terms
of $f_B$. A failure in this simple situation would
raise very serious doubts about the validity or at least
the practical relevance
of the $1/m_Q$ expansion for treating even fully
inclusive nonleptonic transitions; this would leave only
semileptonic transitions in the domain of their applicability.
Such a breakdown of quark-hadron duality would
{\em a priori} appear
as a quite conceivable and merely disappointing
scenario. However such an
outcome would have to be seen as quite puzzling
{\em a posteriori}; for in our
analysis we have not discerned any sign indicating the existence
of such a fundamental problem or a qualitative distinction between
nonleptonic and semileptonic decays \cite{BLOKMANNEL,OPTICAL}.
Thus even a failure would
teach us a valuable, albeit sad lesson about the intricacies of the
strong interactions;  for the heavy quark expansion is
directly and unequivocally based on QCD with the only additional
assumption concerning the workings of
quark-hadron duality!

The theoretical analysis of the lifetimes of heavy-flavour hadrons
can be improved, refined and extended:

\noindent $\bullet$ {\em improved} by a better
understanding of
quark-hadron duality, preferably by deriving it from QCD or
at least by analysing how it operates in the lepton and photon
spectra of semileptonic and radiative $B$ decays, respectively;

\noindent $\bullet$ {\em refined} by a reliable determination of
in particular, but not only, the baryonic expectation values of
the relevant dimension-six operators;

\noindent $\bullet$ {\em extended} by treating $\Xi _b$ (and
$\Omega _b$) decays.

The extension will certainly have been made very soon;
the refinement should be obtained in the
foreseeable future, probably
from a lattice simulation of QCD; when finally
quark-hadron duality will be derived from QCD cannot be
predicted, of course.

There is a corresponding list of future measurements that are
most likely to probe and advance our understanding:

\noindent $\bullet$ measure $\tau (\Lambda _b)$,
 $\tau (\Xi ^- _b)$ and $\tau (\Xi ^0 _b)$
{\em separately} with good accuracy;

\noindent $\bullet$ confirm $\tau (B_d)\simeq
\bar \tau (B_s)$ within an accuracy of very few percent;

\noindent $\bullet$ verify that $\tau (B^+)$ {\em exceeds}
$\tau (B_d)$ by a few to several percent;

\noindent $\bullet$ in charm decays determine the lifetimes
of $\Xi _c^{0,+}$ and $\Omega _c$ baryons with a few percent
accuracy and measure $\tau (D_s)/\tau (D^0)$ with 1\%
precision.

\noindent {\bf Acknowledgements:} I am deeply indebted to
my collaborators B. Blok, M. Shifman, N. Uraltsev
and A. Vainshtein for sharing their insights with me.
Prodding from many experimental colleagues, in
particular S. Paul and V. Sharma, has been instrumental in
making
me face reality (I think). I have been fortified by many excellent
chinese dinners prepared late at night by Bai-Ju's in South Bend.

\begin{Large}
\begin{bf}
Figure Captions
\end{bf}
\end{Large}

Figure 1: WA diagram for $D^0$ decays with gluon emission.

Figure 2: Diagram yielding the operator $\bar QQ$ upon integrating
out the intermediate two fermion and one antifermion lines.

Figure 3: Diagram yielding the chromomagnetic operator
$\bar Q i\sG Q$.

Figure 4a: Cutting a quark line in Fig.2 to generate WA.

Figure 4b: Cutting an antiquark line in Fig.2 to obtain PI.

Figure 5: Forward scattering amplitude involving WA with its three
poles.

\end{document}